\documentclass[iop]{emulateapj}
\newcommand{\gps}{\ensuremath{g_{\rm P1}}}
\newcommand{\rps}{\ensuremath{r_{\rm P1}}}
\newcommand{\ips}{\ensuremath{i_{\rm P1}}}
\newcommand{\zps}{\ensuremath{z_{\rm P1}}}
\newcommand{\yps}{\ensuremath{y_{\rm P1}}}

\newcommand{\grizy}{\gps\rps\ips\zps\yps}

\def\ra#1#2#3{#1$^{\rm h}$#2$^{\rm m}$#3$^{\rm s}$}
\def\dec#1#2#3{$#1^\circ#2'#3''$}
\def\kms\,{km~s$^{-1}$} 
\def\lesssim{\mathrel{\hbox{\rlap{\hbox{\lower4pt\hbox{$\sim$}}}\hbox{$<$}}}}
\def\gtrsim{\mathrel{\hbox{\rlap{\hbox{\lower4pt\hbox{$\sim$}}}\hbox{$>$}}}}

\usepackage{epsfig,graphicx}
\usepackage{amsmath}
\usepackage{natbib}
\usepackage{comment}
\usepackage{url}
\usepackage[usenames,dvipsnames]{color}

\begin{document}

\title{PS1-14bj: A Hydrogen-Poor Superluminous Supernova With a Long Rise and Slow Decay}
\submitted{ApJ in press}
\email{rlunnan@astro.caltech.edu}

\def\cit{1}
\def\cfa{2}
\def\ou{3}
\def\jhu{4}
\def\stsci{5}
\def\ua{6}
\def\okc{7}
\def\nyu{8}
\def\ucsc{9}
\def\ila{10}
\def\ilp{11}
\def\lco{12}
\def\gem{13}
\def\uch{14}
\def\qub{15}
\def\ifa{16}
\def\dur{17}
\def\ein{18}

\author{R.~Lunnan\altaffilmark{\cit,\cfa},
 R.~Chornock\altaffilmark{\ou}, 
 E.~Berger\altaffilmark{\cfa},
 D.~Milisavljevic\altaffilmark{\cfa},
 D.~O.~Jones\altaffilmark{\jhu}, 
 A.~Rest\altaffilmark{\stsci},
 W.~Fong\altaffilmark{\ua,\ein},
 C.~Fransson\altaffilmark{\okc},
 R.~Margutti\altaffilmark{\nyu},
 M.~R.~Drout\altaffilmark{\cfa},
 P.~K.~Blanchard\altaffilmark{\cfa},
 P.~Challis\altaffilmark{\cfa},
 P.~S.~Cowperthwaite\altaffilmark{\cfa},
 R.~J.~Foley\altaffilmark{\ucsc,\ila,\ilp}
 R.~P.~Kirshner\altaffilmark{\cfa},
 N.~Morrell\altaffilmark{\lco},
 A.~G.~Riess\altaffilmark{\stsci,\jhu}
 K.~C.~Roth\altaffilmark{\gem},
 D.~Scolnic\altaffilmark{\uch},
 S.~J.~Smartt\altaffilmark{\qub},
 K.~W.~Smith\altaffilmark{\qub},
 V.~A.~Villar\altaffilmark{\cfa},
 K.~C.~Chambers\altaffilmark{\ifa},
 P. W. Draper\altaffilmark{\dur}
 M.~E.~Huber\altaffilmark{\ifa},
 N.~Kaiser\altaffilmark{\ifa},
 R.-P. Kudritzki\altaffilmark{\ifa}, 
 E.~A.~Magnier\altaffilmark{\ifa}
 N.~Metcalfe\altaffilmark{\dur}, and
 C.~Waters\altaffilmark{\ifa}
   }

\altaffiltext{\cit}{Department of Astronomy, California Institute of Technology, 1200 East California Boulevard, Pasadena, CA 91125, USA}
\altaffiltext{\cfa}{Harvard-Smithsonian Center for Astrophysics, 60 Garden St., Cambridge, MA 02138, USA}
\altaffiltext{\ou}{Astrophysical Institute, Department of Physics and Astronomy, 251B Clippinger Lab, Ohio University, Athens, OH 45701, USA}
\altaffiltext{\jhu}{Department of Physics and Astronomy, Johns Hopkins University, 3400 North Charles Street, Baltimore, MD 21218, USA}
\altaffiltext{\stsci}{Space Telescope Science Institute, 3700 San Martin Drive, Baltimore, MD 21218, USA}
\altaffiltext{\ua}{Steward Observatory, University of Arizona, 933 North Cherry Avenue, Tucson, AZ 85721, USA}
\altaffiltext{\okc}{The Oskar Klein Centre \& Department of Astronomy, Stockholm University, AlbaNova, SE-106 91 Stockholm, Sweden}
\altaffiltext{\nyu}{New York University, Physics department, 4 Washington Place, New York, NY 10003, USA}
\altaffiltext{\ucsc}{Department of Astronomy and Astrophysics, University of California, Santa Cruz, CA 95064, USA}
\altaffiltext{\ila}{Astronomy Department, University of Illinois at Urbana-Champaign, 1002 W. Green Street, Urbana, IL 61801, USA}
\altaffiltext{\ilp}{Department of Physics, University of Illinois at Urbana-Champaign, 1110 W. Green Street, Urbana, IL 61801, USA}
\altaffiltext{\lco}{Las Campanas Observatory, Carnegie Observatories, Casilla 601, La Serena, Chile}
\altaffiltext{\gem}{Gemini Observatory, 670 North Aohoku Place, Hilo, HI 96720, USA}
\altaffiltext{\uch}{Department of Physics, The University of Chicago, Chicago,IL 60637}
\altaffiltext{\qub}{Astrophysics Research Centre, School of Mathematics and Physics, Queen's University Belfast, Belfast BT7 1NN, UK}
\altaffiltext{\ifa}{Institute for Astronomy, University of Hawaii at Manoa, Honolulu, HI 96822, USA}
\altaffiltext{\dur}{Department of Physics, Durham University, South Road, Durham DH1 3LE, UK} 
\altaffiltext{\ein}{Einstein Fellow}

\begin{abstract}
We present photometry and spectroscopy of PS1-14bj, a hydrogen-poor superluminous supernova (SLSN) at redshift $z=0.5215$ discovered in the last months of the Pan-STARRS1 Medium Deep Survey. PS1-14bj stands out by its extremely slow evolution, with an observed rise of $\gtrsim 125$ rest-frame days, and exponential decline out to $\sim 250$ days past peak at a measured rate of $0.01~{\rm mag~day}^{-1}$, consistent with fully-trapped $^{56}$Co decay. This is the longest rise time measured in a SLSN to date, and the first SLSN to show a rise time consistent with pair-instability supernova (PISN) models. Compared to other slowly-evolving SLSNe, it is spectroscopically similar to the prototype SN\,2007bi at maximum light, though lower in luminosity ($L_{\rm peak} \simeq 4.6 \times 10^{43} {\rm erg s}^{-1}$) and with a flatter peak than previous events. PS1-14bj shows a number of peculiar properties, including a near-constant color temperature for $>200$ days past peak, and strong emission lines from [\ion{O}{3}] $\lambda$5007 and [\ion{O}{3}] $\lambda$4363 with a velocity width of $\sim$3400 \kms\, in its late-time spectra. These both suggest there is a sustained source of heating over very long timescales, and are incompatible with a simple $^{56}$Ni-powered/PISN interpretation. A modified magnetar model including emission leakage at late times can reproduce the light curve, in which case the blue continuum and [\ion{O}{3}] features are interpreted as material heated and ionized by the inner pulsar wind nebula becoming visible at late times. Alternatively, the late-time heating could be due to interaction with a shell of H-poor circumstellar material.
\end{abstract}


\section{Introduction}
Superluminous supernovae (SLSNe), characterized by peak luminosities 10-100 times higher than ordinary core-collapse and Type Ia SNe, are a rare type of transients discovered in untargeted surveys over the past decade. Their extreme energetics cannot be explained by the same mechanism as ordinary SNe, requiring either an additional energy source beyond $^{56}$Ni decay, or an exotic explosion mechanism. 
From an observational point of view, SLSNe can be classified into at least two distinct categories \citep{gal12}: H-rich SLSNe (`SLSN-II') show Balmer lines in their spectra, and most are similar to Type IIn SNe and likely indicating strong interaction with circumstellar material (CSM) (e.g., \citealt{ock+07,slf+07,scs+10,rfg+11}). In this case, a large fraction of the kinetic energy of the ejecta can be converted into radiation, powering the large luminosities. 

In contrast, H-poor SLSNe (`SLSN-I') typically have spectra distinct from any other known SN types, characterized by a very blue continuum with a few, broad features from intermediate-mass elements (e.g., \citealt{qkk+11, ccs+11,msp+16}). The primary power source for this class is debated: CSM interaction can explain the energetics, but would require several solar masses of H-poor material shed in the last few years to decades before explosion \citep{ci11,gb12,cw12b,mbt+13}. An alternative scenario is energy injection from a central engine, such as the spin-down of a newborn magnetar \citep{kb10,woo10,dhw+12}. $^{56}$Ni decay, which powers the optical light curve of ordinary type Ibc SNe, can generally be ruled out for this subclass as the fast timescales at peak are incompatible with the large Ni masses that would be required to power the luminosities (e.g., \citealt{qkk+11,ccs+11,lcb+13,isj+13}).

However, some H-poor SLSNe do show slow decay rates that match that of radioactive $^{56}$Co. \citet{gal12} proposed that these be grouped into a third subclass, `SLSN-R' (the `R' stands for radioactivity), modeled on the unusual H-poor SLSN SN\,2007bi \citep{gmo+09, ysv+10}. SN\,2007bi had both a slow decay rate and a nebular phase spectrum consistent with several solar masses of $^{56}$Ni produced in the explosion, leading \citet{gmo+09} to propose that SN\,2007bi was a pair-instability supernova (PISN). This mechanism is theorized to be the final fate of stars with initial masses  $140~M_{\odot} \lesssim {\rm M} \lesssim 250~M_{\odot}$, where the high temperatures in the oxygen core leads to electron-positron pair production triggering a nuclear explosion \citep{brs67}. While originally thought to be mainly relevant for Population III stars in the early universe \citep{wbh07}, models including rotation are able to produce PISNe also from non-zero metallicity stars \citep{cw12a,yhm+13,mlp+16}.

The interpretation of SN\,2007bi as a PISN (and by extension, the interpretation of `SLSN-R' as a separate class from other H-poor SLSNe and powered by radioactivity) is controversial, however. A key prediction of PISN models is a long ($\gtrsim 100~{\rm days}$) rise-time due to the large ejecta mass, which in turn leads to a long diffusion time through the ejecta. As SN\,2007bi was discovered near peak, the rise was not observed, and CSM interaction or magnetar spin-down models can also fit the observed light curve. In addition, other H-poor SLSNe with decay rates and spectroscopic properties similar to SN\,2007bi have since been discovered, showing shorter rise times incompatible with PISN models but well-fit by magnetar spin-down models \citep{nsj+13}, thus casting doubt also on the nature of SN\,2007bi. Finding more slowly-evolving SLSNe with well-sampled light curves is therefore crucial in shedding light on the origins of this potential subclass.

Here, we present the discovery and analysis of PS1-14bj, a slowly evolving H-poor SLSN discovered by the Pan-STARRS Medium Deep Survey (PS1/MDS). The photometry and spectroscopic observations are presented in Section~\ref{sec:data}. We analyze the light curves, including the color evolution and blackbody fits, and use this data to construct a bolometric light curve and estimate the total radiated energy in Section~\ref{sec:lc}. Our spectroscopic sequence, ranging from $-51$ to $+202$ days relative to peak light, is discussed in Section~\ref{sec:spectra}. We discuss the host galaxy properties in Section~\ref{sec:hostgal}. Section~\ref{sec:models} discusses our observations in the context of different models for powering SLSNe, and we summarize our conclusions in Section~\ref{sec:conc}. All calculations in this paper assume a $\Lambda$CDM cosmology with $H_0 = 70$~km~s$^{-1}$~Mpc$^{-1}$, $\Omega_{\rm M} = 0.27$ and $\Omega_{\Lambda} = 0.73$ \citep{ksd+11}.

\section{Observations}
\label{sec:data}

\subsection{Pan-STARRS1 Summary and Photometry}

The PS1 telescope on Haleakala is a high-\'{e}tendue wide-field survey
instrument, with a 1.8 m diameter primary mirror and a $3.3^\circ$
diameter field of view, imaged by an array of sixty $4800\times 4800$
pixel detectors with a pixel scale of $0.258''$
\citep{PS1_system,PS1_GPCA}.  The \grizy broadband filters and photometric system are described in detail in \citet{tsl+12}. 

PS1/MDS operated from late 2009 to early 2014. PS1/MDS consists of 10 fields (each with a
single PS1 imager footprint) observed in \gps\rps\ips\zps with a typical cadence of 3~d in each filter, to a typical nightly depth of $\sim 23.3$ mag ($5\sigma$); \yps is used near full moon with a typical depth of $\sim 21.7$ mag.
The standard reduction, astrometric solution, and stacking of the
nightly images are done by the Pan-STARRS1 Image Processing Pipeline (IPP) system \citep{PS1_IPP, PS1_astrometry} on a computer cluster at the Maui High Performance Computer Center. For the transients search, the nightly MDS stacks were transferred
to the Harvard FAS Research Computing cluster, where they were
processed through a frame subtraction analysis using the {\tt photpipe}
pipeline developed for the SuperMACHO and ESSENCE surveys \citep{rsb+05, gsc+07, mpr+07,rsf+14}.

PS1-14bj was first detected in PS1/MDS imaging on 2013 November 22.6 (UT dates are used throughout this paper), at coordinates RA = \ra{10}{02}{08.433}, Dec = \dec{+03}{39}{19.02} (J2000). The detection image was also the first image taken of this field after 2013 May 10.3, so we do not have any recent limits to constrain the explosion date further. The PS1/MDS light curve shows an unusually slow rise over an observed time scale of two months, and PS1-14bj was selected for follow-up spectroscopy. An initial MMT spectrum taken on 2014 March 8.3 proved inconclusive, so an additional spectrum was obtained with Gemini-North on 2014 March 28.4.  The broad spectral features in those data show an excellent match with the earliest spectrum of the unusual H-poor SLSN SN\,2007bi \citep{gmo+09,ysv+10}, and an approximate redshift of $z \approx 0.54$ was determined by cross-correlating the SN spectrum with that of SN\,2007bi. As the supernova faded, weak narrow-line [\ion{O}{3}] $\lambda$5007 and H$\beta$ emission from the host galaxy became visible, yielding a precise redshift of $z = 0.5215$, which is adopted throughout this paper.  The somewhat higher redshift estimate obtained by comparison with SN\,2007bi is due to the higher expansion velocity of that object compared to PS1-14bj (Section~\ref{sec:spectra}). We note that given this redshift, the observed peak magnitude of PS1-14bj is $-20.75~{\rm mag}$ (i-band). This is fainter than the proposed SLSN cutoff at $-21~{\rm mag}$ in \citet{gal12}, illustrating that such a cutoff is arbitrary.

We determine the time of peak light by fitting a low-order polynomial to our constructed bolometric light curve (Section~\ref{sec:bollc}), yielding a best-fit time of peak of MJD 56801.3 $\pm 5~{\rm days}$. All phases reported are relative to the time of peak light. 

\subsection{Additional Photometry}
As PS1-14bj was discovered at the very end of PS1/MDS, we have continued to follow it up from other telescopes to continue the light curve. Additional optical imaging was mainly obtained with MMTCam\footnote{\url{https://www.cfa.harvard.edu/mmti/wfs.html}}, an f/5 imager on the 6.5~m MMT telescope; with the Low Dispersion Survey Spectrograph (LDSS3) and the Inamori-Magellan Advanced Camera for Surveys (IMACS; \citealt{dhb+06}) on the 6.5~m Magellan telescopes; with the Gemini Multi-Object Spectrograph (GMOS; \citealt{hja+04}) on the 8~m Gemini Telescopes as part of acquisition for spectroscopy; and some additional images from the Large Binocular Cameras on the 8.4~m Large Binocular Telescope \citep{sdg+08}, the RETRactable Optical CAmera for Monitoring (RETROCAM; \citealt{mbd+05}) on the MDM 2.4~m Hiltner telescope, and the 1.3~m McGraw-Hill telescope at MDM Observatory in direct imaging mode with the R4K detector. We also obtained two epochs of NIR imaging with the FourStar Infrared Camera on the 6.5~m Magellan/Baade telescope \citep{pbb+08}. 

All optical images were bias-subtracted, flat-fielded and stacked using standard procedures in IRAF\footnote{IRAF is
  distributed by the National Optical Astronomy Observatory,
    which is operated by the Association of Universities for Research
    in Astronomy, Inc., under cooperative agreement with the National
    Science Foundation.}. To correct for the underlying host galaxy flux, we used the ISIS subtraction package \citep{al98} to subtract off the PS1/MDS pre-explosion template images (Section~\ref{sec:hostgal_phot}) before measuring the SN flux in the subtracted images using aperture photometry. 
    
The field of view and consequently the stars available for photometric calibration varies between the instruments we used to follow PS1-14bj. To ensure consistency, we first measured the light curve relative to two nearby stars that are present in all images. The absolute calibration was determined from a set of IMACS images, comparing photometry of $> 10$ stars in each filter to SDSS. All photometry has been corrected for foreground extinction according to \citet{sf11} ($E(B-V) = 0.019~{\rm mag}$), and is listed in Table~\ref{tab:phot}. The light curves are plotted in Figure~\ref{fig:obslc}. 
    
The FourStar images were calibrated, aligned and co-added using the IRAF/{\tt FSRED} package (A. Monson 2013, private communication). As we lack pre-explosion host galaxy templates for subtraction in the NIR, we instead use the SED fit to the host galaxy photometry (Section~\ref{sec:hostgal}) to estimate the expected host galaxy flux. We determine the total flux using aperture photometry, with the zeropoint determined from 2MASS stars in the field, and subtract the expected host contribution numerically. The resulting $J$-band magnitudes are listed in Table~\ref{tab:phot}. $K_S$-band imaging was also performed in the second epoch, but the detection is marginal and close to the expected host galaxy flux within the uncertainties. Image subtraction is needed to determine the SN contribution, and we therefore do not use the $K_S$-band data in further analysis.

\begin{figure}
\centering
\includegraphics[width=3.5in]{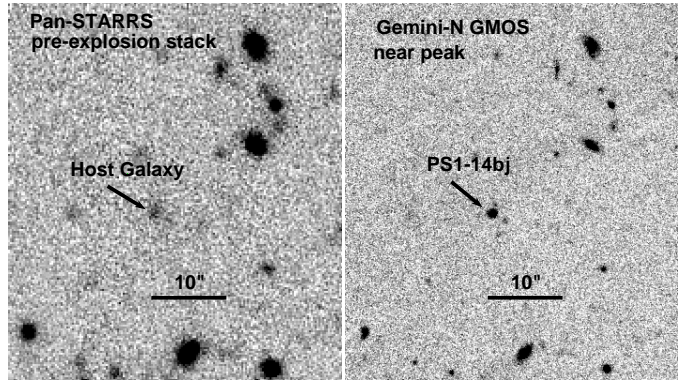}
\caption{Stacked \ips\ PS1/MDS pre-explosion image of the field around PS1-14bj (left), compared to an $i$-band image from GMOS taken near peak (right). A faint host galaxy is clearly seen at the supernova position. In the GMOS image, which has  significantly better seeing (0.4\arcsec\, FWHM, compared to 1.3\arcsec\, in the template), it appears that the host galaxy may have some structure or multiple components. 
\label{fig:hostgal}}
\end{figure}

\begin{figure*}
\centering
\includegraphics[width=7in]{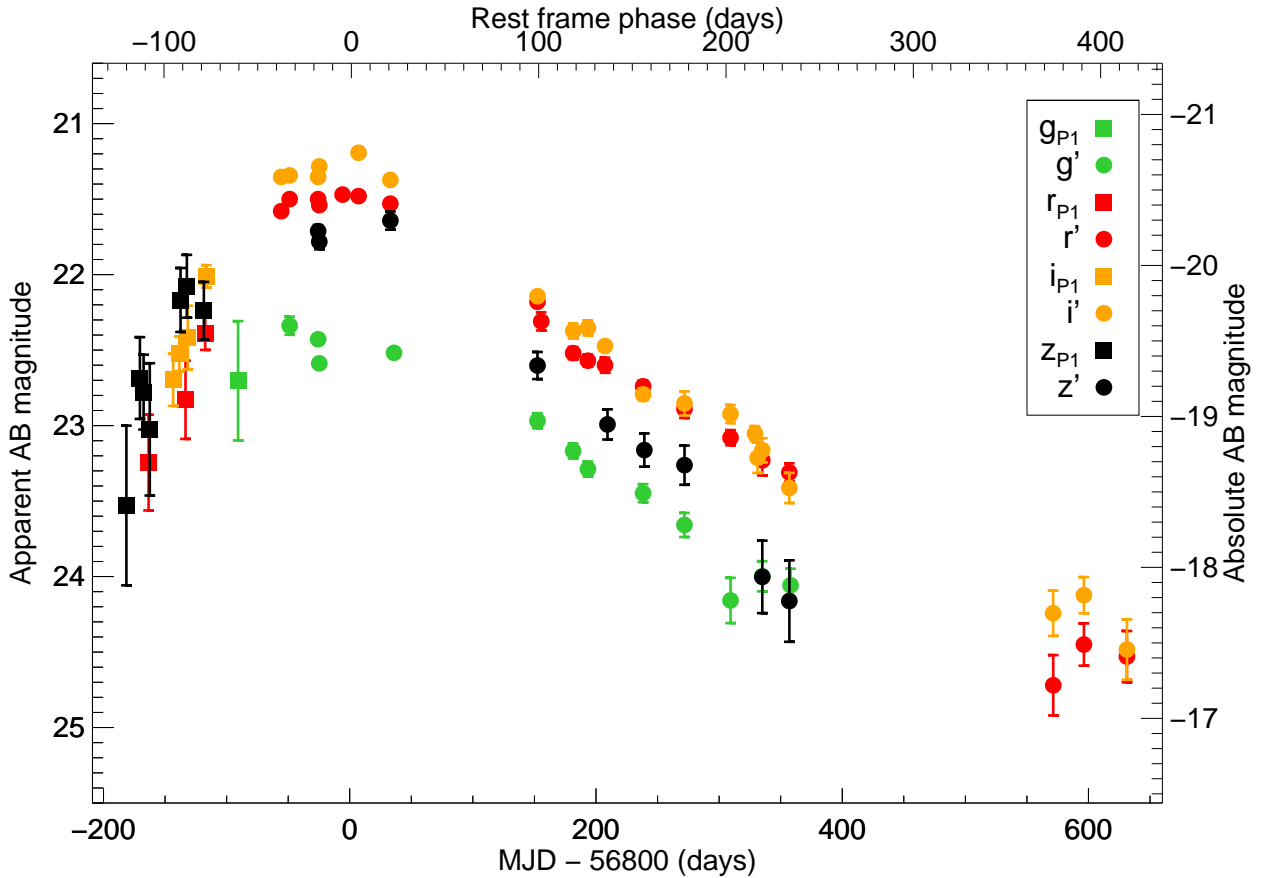}
\caption{Observed light curve of PS1-14bj. The first 2.5 months of \rps\ips\zps\ data are from PS1/MDS, while the remaining photometry is from a variety of telescopes as detailed in Table~\ref{tab:phot}. PS1-14bj was spectroscopically classified after PS1/MDS had ceased taking data and so there is a gap in coverage on the rise; there are also two gaps on the decline due to PS1-14bj being in solar conjunction. Note the long observed time scales. 
\label{fig:obslc} }
\end{figure*}

\subsection{Spectroscopy}
 We obtained 11 spectra of PS-14bj between 2014 March 8 and 2015 March 28 using the Blue Channel spectrograph on the MMT \citep{swf89}, IMACS and LDSS3C at Magellan, and both GMOS-N and GMOS-S on the 8~m Gemini telescopes.  
  We use IRAF to perform bias subtraction, flat fielding, and spectral extraction.  We use observations of spectrophotometric standard stars (archival in the case of GMOS) to apply a flux calibration and correct for telluric absorption.
  The final GMOS-N observation was taken in nod-and-shuffle mode \citep{gb01}, but we find that the sky subtraction is adequate without performing the subtraction of nod pairs, so we simply shift and combine all exposures with the same grating tilt prior to spectral extraction.
Observations taken in red-sensitive setups used order-blocking filters. The slits were generally aligned near the parallactic angle \citep{fil82}, except at low airmass or with IMACS, which has an atmospheric dispersion compensator, so the relative flux scales should be reliable, as demonstrated below by comparison to the photometry in Section~\ref{sec:bb}.
   In some cases, we combine spectra taken on consecutive nights or in setups with complementary wavelength coverage.  A summary of all spectroscopic observations is given in Table~\ref{tab:spec} and the spectra are shown in Figure~\ref{fig:spec_montage}.

\begin{figure}
\centering
\includegraphics[width=3.5in]{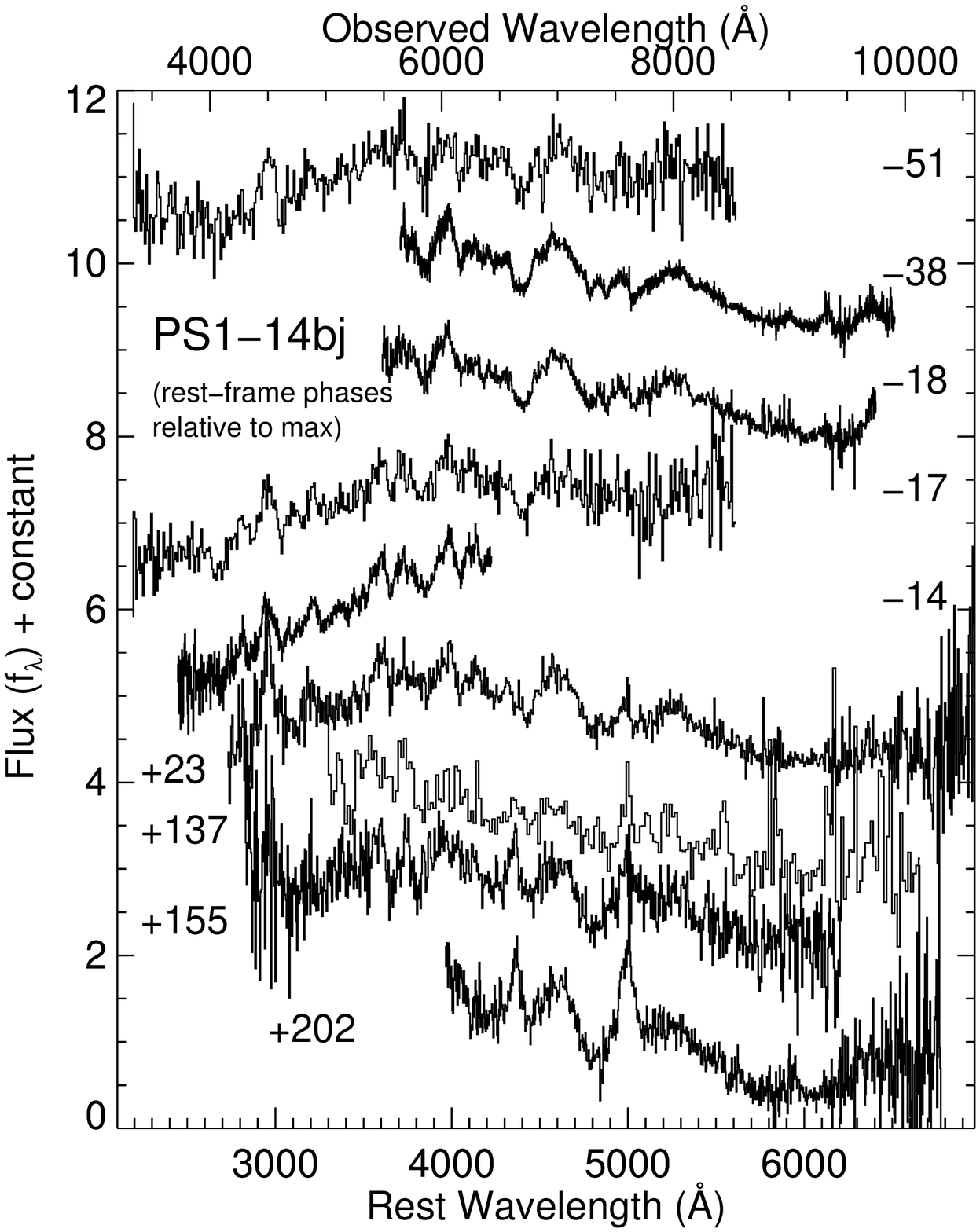}
\caption{Spectroscopic sequence of PS1-14bj, labeled by the rest-frame phases relative to bolometric maximum light.  The spectra have been binned for display purposes.
\label{fig:spec_montage}}
\end{figure}

\subsection{Host Galaxy Photometry}
\label{sec:hostgal_phot}
To search for a host galaxy, we stack the PS1/MDS data from the four observing seasons prior to the detection of PS1-14bj. No host is detected in the ``standard'' deep stacks (consisting of the 80 observations with best image quality), but by stacking all the available data (about 200 individual observations/filter) we are able to go significantly deeper at the expense of worse FWHM in the resulting stacks. A faint galaxy is detected at the position of PS1-14bj in all filters; as an example, the \ips\ stack is shown in Figure~\ref{fig:hostgal}. Photometry in a 2\arcsec aperture yields $\gps = 24.57 \pm 0.15$~mag, $\rps = 23.83 \pm 0.13$~mag, $\ips = 23.93 \pm 0.15$~mag, and $\zps = 23.51 \pm 0.16$~mag. We use these deep stacks as subtraction templates for the non-PS1 photometry (the PS1 photometry is already host subtracted as part of the pipeline). The properties of the host galaxy itself are discussed in Section~\ref{sec:hostgal}.

\subsection{Comparison Data: LSQ14an}
\label{sec:lsq14an}
LSQ14an is a supernova discovered by the La Silla Quest survey in early 2014, at coordinates RA = \ra{12}{53}{47.83}, Dec = \dec{-29}{31}{27.2}. It was classified by PESSTO \citep{pessto} as a post-peak 2007bi-like SLSN at redshift $z = 0.163$ \citep{lgf+14}. We have obtained a spectroscopic sequence of LSQ14an, and find it to be a remarkably close match to PS1-14bj (Section~\ref{sec:spectra}). The details of our observations of LSQ14an are given in Table~\ref{tab:lsqspec}. In addition, $griz$ photometry was obtained with each spectrum; the $g$- and $r$-band light curves are shown in Figure~\ref{fig:lccomp}. We lack supernova-free galaxy templates in our specific camera and filter combinations, and therefore the image subtraction method can not yet be applied to analyze the light curves in detail. However, our data demonstrates that LSQ14an was indeed discovered on the decline (as was inferred from the initial PESSTO spectrum), and furthermore that it is also a SLSN with a slow decline rate, establishing it as a reasonable comparison object for PS1-14bj. The host galaxy is visible in the Pan-STARRS 3$\pi$ image stack, at $g_{\rm P1}= 20.93\pm 0.11$ and $r_{\rm P1}= 20.50 \pm 0.12$ (\citealt{che15}; Inserra et al. in prep) indicating that the flattening of the lightcurve in Figure 4 is almost certainly due
to contamination of the host.

The spectra of LSQ14an are used as a basis for comparison in Section~\ref{sec:spectra}. Full spectral analysis and further consideration of the light curve await proper host subtraction after the SN fades, which we defer to future work.

\section{Light Curve and Energetics}
\label{sec:lc}

\subsection{Light Curve Comparisons}
The $r$- and $z$-band light curves of PS1-14bj are shown in Figure~\ref{fig:lccomp}, compared to other slowly-evolving H-poor SLSNe from the literature: SN\,2007bi, PTF12dam, iPTF13ehe and PS1-11ap. We also show our own light curve of LSQ14an, establishing it as another slowly-evolving object. All light curves have been corrected for cosmological expansion following

\begin{equation}
M = m - 5\log(D_L/10~{\rm pc}) + 2.5 \log(1+z)
\end{equation}

\noindent \citep{hbb+02}, where $m$ is the apparent AB magnitude and $D_L$ is the luminosity distance. This is not a full $K$-correction; filters have been chosen to correspond to approximately similar rest wavelengths as indicated in the legends in order to facilitate comparison. In both filters, the light curve of PS1-14bj is significantly broader than the other events, in particular having a significantly longer rise time than any of the other events, and a flatter light curve with a lower peak luminosity. In the bluer band, the decline of PS1-14bj appears shallower than the other events, but the decline in the redder filter is similar to the other events. The decay slope of LSQ14an appears to be even shallower than PS1-14bj, but the photometry plotted for this event includes an unknown contribution from the host galaxy (Section~\ref{sec:lsq14an}). 

Interestingly, the light curve decline is not entirely smooth particularly in $i$- and $z$-band (Figure~\ref{fig:obslc}), but shows some undulations around 100-200~days. The light curve of SN\,2007bi shows a similar kink \citep{gmo+09,ysv+10}, and recently \citet{nbs+16} showed that the well-studied SN\,2015bn also displays a ``knee'' in the light curve decline. Such deviations from a smooth decline may therefore not be unusual in slowly-evolving SLSNe.

\begin{figure*}
\centering
\begin{tabular}{cc}
\includegraphics[width=3.4in]{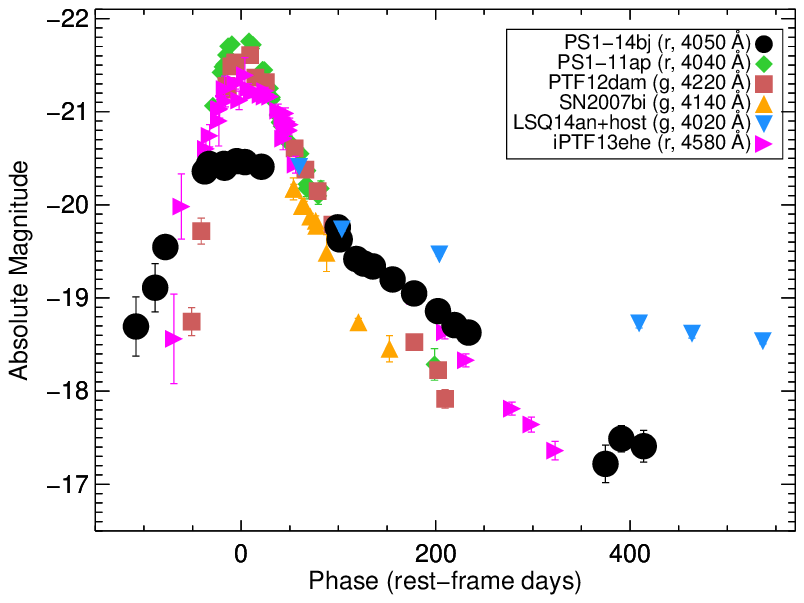} & \includegraphics[width=3.4in]{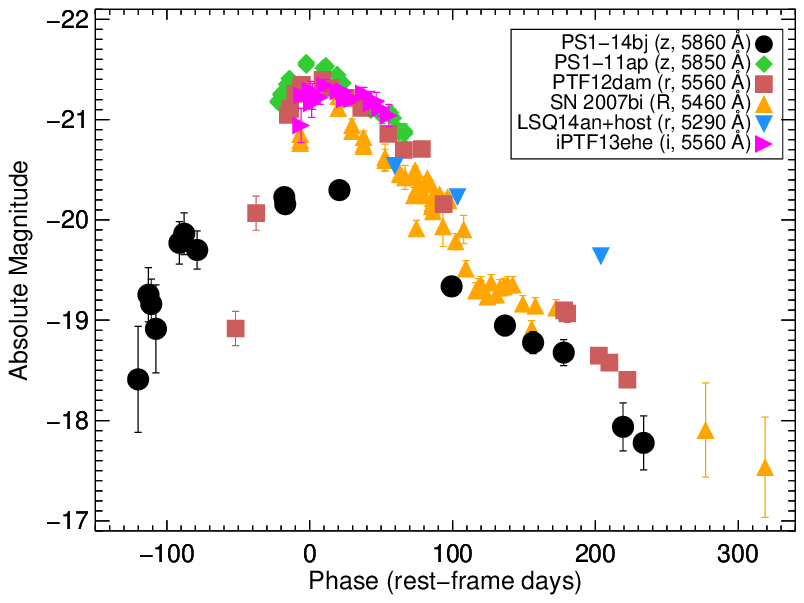}
\end{tabular}
\caption{Light curve of PS1-14bj at effective wavelengths 4000\,\AA\, ($r$-band, left) and 5800\,\AA\, ($z$-band, right), compared to other ``slowly-evolving'' H-poor SLSNe: SN\,2007bi \citep{gmo+09,ysv+10}, PS1-11ap \citep{nsj+13,msk+14}, PTF12dam \citep{nsj+13} and iPTF13ehe \citep{yqo+15}. PS1-14bj and PS1-11ap are both at redshift $z\sim 0.5$ while SN\,2007bi and PTF12dam are at redshift $z\sim 0.1$; filters plotted are chosen to correspond to approximately the same rest wavelengths, and have been corrected for cosmological expansion. PS1-14bj shows a notably slower rise and flatter peak than PS1-11ap and PTF12dam, and a significantly shallower decline at a rest wavelength of $\sim 4000$\,\AA\, than the other objects. At a rest wavelength of $\sim 5800$\,\AA, the decline rate of PS1-14bj is similar to the other slowly-evolving SLSNe, but still shows a significantly broader light curve and lower peak luminosity.  The LSQ14an data have not had the host contribution subtracted due to the current lack of templates, but demonstrate that it is also has a slow decline rate.
\label{fig:lccomp}}
\end{figure*}

\subsection{Color Evolution and Blackbody Fits}
\label{sec:bb}

The observed color evolution of PS1-14bj is shown in Figure~\ref{fig:color}. The color evolution is almost flat over the $> 300$~days covered by our light curve data, and shows an overall trend towards \textit{bluer} colors with time in $i-z$. This is different from the color evolution observed so far in virtually all H-poor SLSNe, which usually start out very blue but turn redder as the ejecta expand and cool (see e.g. PS1-11ap, shown in Fig.~\ref{fig:color} for a more typical example). The near-constant color of PS1-14bj suggests there must be a sustained source of heating over very long timescales. 

\begin{figure}
\centering
\includegraphics{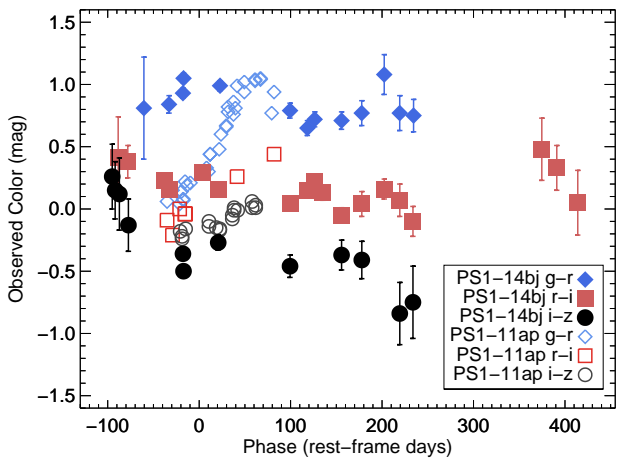}
\caption{Color evolution of PS1-14bj (filled symbols). The open symbols show the more typical H-poor SLSN PS1-11ap \citep{msk+14}, which was at almost exactly the same redshift as PS1-14bj ($z = 0.524$), facilitating comparison. PS1-14bj is significantly redder than PS1-11ap in its early phases, particularly seen in the observed $g-r$ color. Also note the remarkably flat color evolution of PS1-14bj, with an overall trend towards bluer colors with time, particularly in $i-z$. In contrast, PS1-11ap (like most supernovae) turns redder with time. }
\label{fig:color}
\end{figure}

We fit blackbody curves to the SED of PS1-14bj in order to track the temperature and radius of the photosphere with time, using the IDL {\tt mpfitfun}\footnote{\url{http://cow.physics.wisc.edu/~craigm/idl/idl.html}} framework to fit a Planck function to the observed photometry. Two example SEDs are shown in Figure~\ref{fig:sedfit}, for the two epochs where we also have NIR photometry. The spectra closest in time to the NIR photometry are also shown for comparison. The flux in $g$-band is affected by strong absorption, and including $g$-band in the fit leads to a cooler derived blackbody temperature and a larger blackbody radius. 

\begin{figure}
\centering
\includegraphics[width=3.5in]{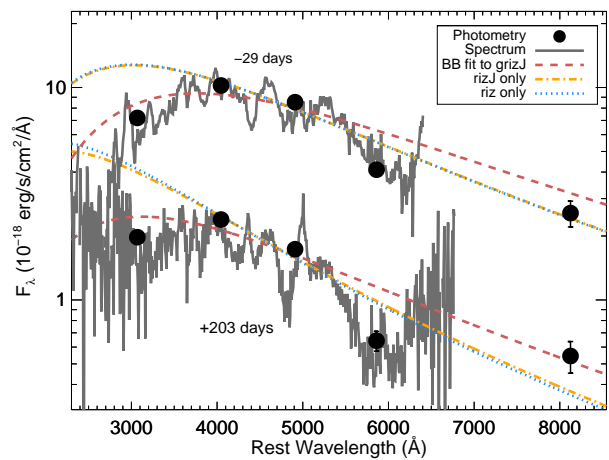}
\caption{Example SED fits for the two epochs where we have $J$-band photometry. The black points show the photometry, while the spectra taken closest in time are shown in gray. The $g$-band flux is suppressed by line blanketing, and including it in the SED fit (red dashed curve) leads to a cooler inferred temperature and larger radius. We therefore exclude $g$-band measurements from our blackbody fits when calculating the bolometric correction in the red, and for most epochs only fit $riz$, here illustrated by the blue dotted curve. Including the $J$-band photometry (orange dot-dashed curve) yields a very similar fit, indicating that the $riz$ fit is a reasonable approximation of the SED shape redwards of the observed bands.}
\label{fig:sedfit}
\end{figure}

Figure~\ref{fig:bb} shows the resulting blackbody temperatures (top) and radii (middle), showing fits both to all filters (black points) and excluding $g$-band (red points). The plotted error bars are the formal uncertainties returned from the blackbody fits given the photometry uncertainties. While the formal uncertainty is small (in many cases $< 500~{\rm K}$ and thus too small to be seen on the plot in Figure~\ref{fig:bb}), the fact that the difference between the set of fits is larger than the calculated error bars indicates that the systematic uncertainty associated with line blanketing is the dominant source of uncertainty on the actual temperature. We note that the trends in temperature and radius are the same for both sets of fits, however.

The temperature fits show the same trends as are seen in the color evolution: the blackbody temperature is remarkably slowly evolving as well, and is in fact increasing with time over the entire time sampled by our light curve. With the overall flux declining, while the temperature is constant or increasing, the best-fit blackbody radius is decreasing as the light curve declines. We note that combining the estimated radius at peak (3--5 $\times 10^{15}~{\rm cm}$) and photospheric velocity (5000~\kms\,; Section~\ref{sec:spectra}) yields a rise-time of 70--115~days, well within our observed rise.

\begin{figure}
\centering
\includegraphics[width=3.4in]{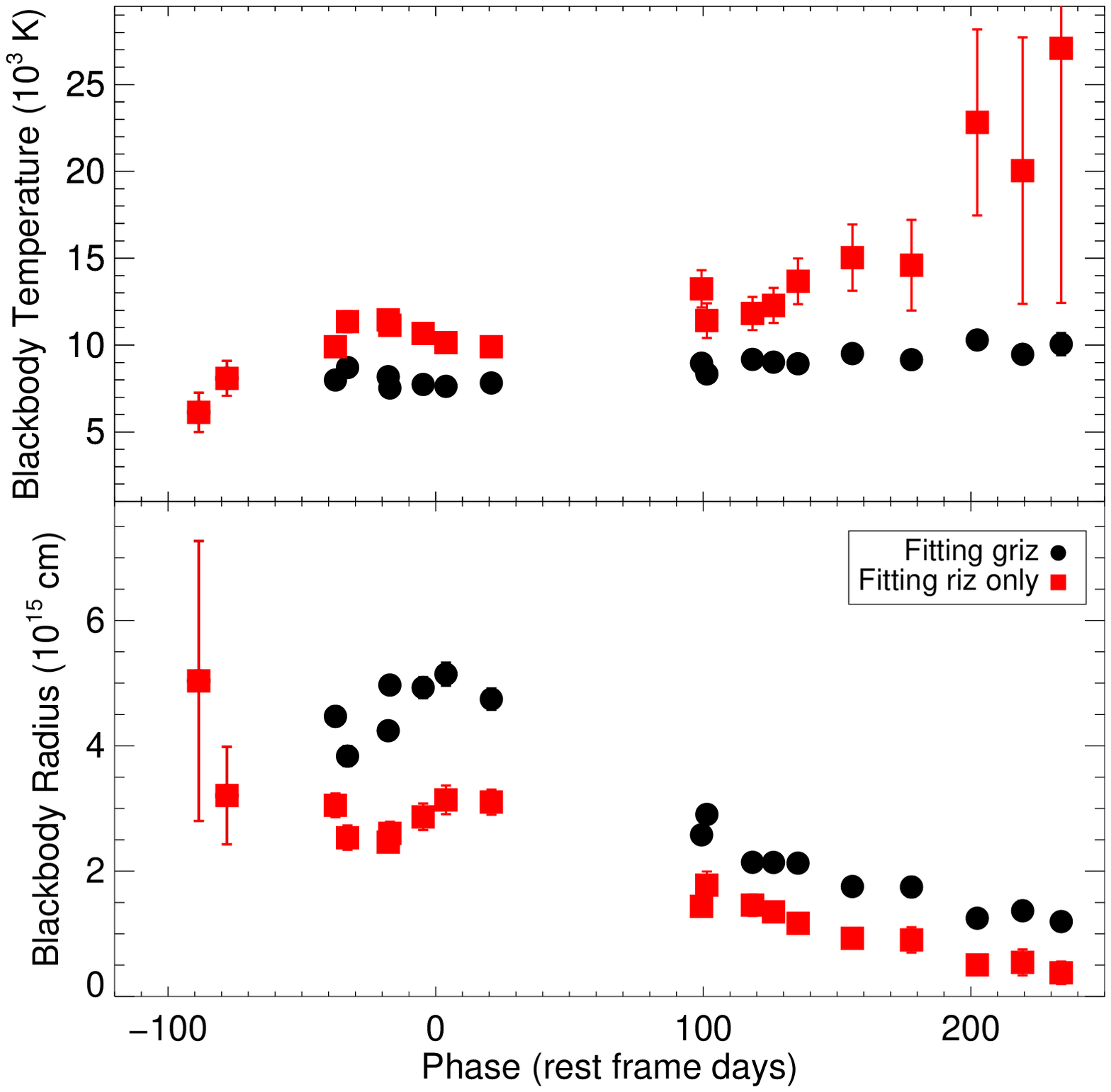}
\caption{Blackbody temperature (top) and radius (bottom) of PS1-14bj. Black points show fits to all photometry, while the red points exclude $g$-band data (since the spectroscopy shows $g$-band is affected by line blanketing; Figure~\ref{fig:sedfit}). The difference between the two sets of fits are generally larger than the derived error bars from the fit, suggesting that these line effects are the largest source of uncertainty in the absolute temperature. The general trends remain the same, however. }
\label{fig:bb}
\end{figure}

\subsection{Bolometric Light Curve and Total Radiated Energy}
\label{sec:bollc}
In order to construct a pseudo-bolometric light curve, we first sum up the observed photometry. In order to account for flux outside of our observed bandpasses, we add to this a blackbody tail redwards of the observed photometry, using the temperatures derived from our blackbody fits. For the purposes of this correction, we use the fits that exclude $g$-band.  Blueward of the observed $g$-band, the spectra show that the flux is suppressed by line blanketing, and, in particular, not well approximated by a blackbody. We therefore do not attempt to add a correction for missing blue flux, and our reported bolometric light curve should be considered a lower limit. We do have a few spectra with coverage extending bluer than $g$-band; from summing the total flux in the 2014 April 28 Blue Channel spectrum bluewards of $g$-band, as well as approximating  the spectral shape in the blue as a polynomial and extrapolating, we find that the missing flux is of order $\lesssim 5 \%$ of the total.

The resulting bolometric light curve is shown in Figure~\ref{fig:bollc}, and the data is listed in Table~\ref{tab:bollc}. Open symbols show the light curve from only combining the observed flux at epochs where we have at least three filters, and the closed symbols include the bolometric correction to the red. For the earliest light curve points, where we lack color information, we assume the same SED shape as the earliest multi-filter epochs. Similarly, for the last three epochs we only have the $r-i$ color, and we estimate the bolometric luminosity by assuming the same bolometric correction as for the last two epochs with multi-color data. The $r-i$ color in our latest epochs is consistent with the previous epochs within the uncertainties, so this is not an unreasonable assumption. As the color is somewhat different between these three epochs, we get different results depending on whether we scale from $r$- or $i$-band; we take the mean value as our best estimate. We caution, however, that at these late times the spectra are dominated by line emission rather than continuum, and so the blackbody assumptions used to derive the bolometric corrections will break down. The error bars of the latest light curve points are therefore likely underestimated.

The peak luminosity is $4.56 \times 10^{43}~{\rm erg s}^{-1}$ or $- 20.45~{\rm mag}$. As can also be glanced from Figure~\ref{fig:lccomp}, PS1-14bj is significantly lower luminosity at peak than other slowly-evolving SLSNe. The early light curve decay (out to $\sim 250$~days) is well fit by exponential decay, with a best-fit slope of $(1.00 \pm 0.05) \times 10^{-2}~{\rm mag~day}^{-1}$. We note that this slope is very close to that of the decay of $^{56}$Co to $^{56}$Fe, which is $9.74 \times 10^{-3}~{\rm mag~ day}^{-1}$, and shown by the dashed gray line in Figure~\ref{fig:bb}. However, while noisy, the latest light curve points (around 400~days) lie above the $^{56}$Co prediction; including these points give a best-fit decay slope of $(9.2 \pm 0.4) \times 10^{-3}~{\rm mag~day}^{-1}$.

\begin{figure}
    \centering
    \includegraphics{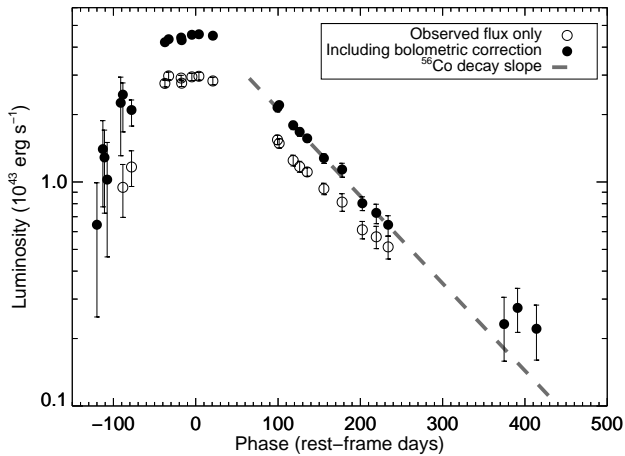}
    \caption{Bolometric light curve of PS1-14bj. The open symbols are created by summing all the flux at epochs where we have at least three filters, a strict lower limit.  The filled symbols correspond to the bolometric light curve including a blackbody correction in the red. The decline luminosity is well fit by exponential decay, with a best-fit slope very close to that of pure Co decay, which is shown by the dashed gray line. }
    \label{fig:bollc}
\end{figure}

We estimate the time of explosion from fitting a $t^2$ polynomial to the rise portion of the light curve, setting the explosion date at MJD 56607.0 (2013 November 11.0) $\pm 6.4~{\rm days}$, with most of the uncertainty being due to the noisy nature of the early light curve. This suggests that PS1-14bj exploded a short time before PS1/MDS resumed observing the field on 2013 November 22.6, so our observed data covers most of the rise. We note that a $t^2$ polynomial is an excellent fit to the rise of the light curve, with a reduced $\chi^2$ value of 1.01. We therefore do not find any evidence for a double-peaked light curve, as was seen in LSQ14bdq \citep{nsj+15b}, and argued to possibly be a ubiquitous feature of H-poor SLSNe \citep{ns16}.
Similarly, we estimate the time of peak by fitting a forth-order polynomial to the data near peak, and find the best-fit time of peak to be MJD 56801.3 $\pm 5~{\rm days}$; here the uncertainty is mainly due to the flatness of the peak. This gives a rise time estimate of 128~days in the rest frame, the longest observed in a SLSN so far. 

Integrating our bolometric light curve over time gives an estimate of the total radiated energy: $E_{\rm rad} \simeq 7.6  \times 10^{50}~{\rm erg}$. This number is comparable to other SLSNe; the lower peak luminosity is compensated by the broad light curve.

\section{Spectroscopic Evolution}
\label{sec:spectra}

To identify the features and estimate the velocities in the photospheric spectra, we use the spectral synthesis code {\tt SYN++} \citep{tnm11}. Figure~\ref{fig:syn++} shows the fit to the combined day $-17$/day $-14$ spectrum, with the main features marked. The gray line is a fit including only \ion{Fe}{2}, \ion{Mg}{2} and \ion{Ca}{2}, which reproduces the main absorption features well. The absorption feature near 3840\,\AA\, is well fit by \ion{Ca}{2}, and the narrow absorptions between 4500--5300\,\AA\, are reasonably fit by \ion{Fe}{2}. The feature near 4390\,\AA\, is usually attributed to \ion{Mg}{2} (e.g., \citealt{gmo+09}), although we find it is best fit by a blend of \ion{Mg}{2} and \ion{Fe}{2}. Including \ion{Co}{2}, \ion{Ni}{2} and \ion{Ti}{2} (red line) helps suppress the flux in the blue, and improves the overall fit. Velocities used in the fit range from 5000--9000~\kms\,, with \ion{Fe}{2} ``detached'' to match the narrow absorption features. We do not find any significant velocity evolution fitting the spectra from day $-38$ to day $+23$. 

\begin{figure}
\centering
\includegraphics[width=3.4in]{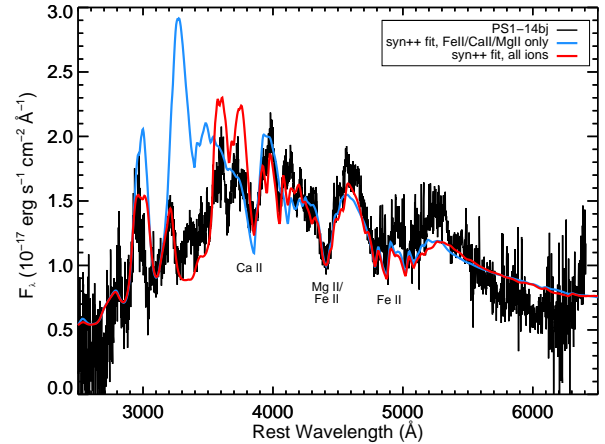}
\caption{{\tt SYN++} fit to the near-peak spectrum of PS1-14bj. The main features are attributed to \ion{Ca}{2}, \ion{Mg}{2} and \ion{Fe}{2} traveling at velocities 5000-9000~\kms\,, as shown in the blue-line fit that only contains these ions. The full fit also includes \ion{Co}{2}, \ion{Ni}{2} and \ion{Ti}{2}, which significantly improves the fit in the blue.}
\label{fig:syn++}
\end{figure}

Figure~\ref{fig:earlyspec} shows a comparison of the early-time spectra of PS1-14bj to three other slowly-evolving H-poor SLSNe: SN\,2007bi, LSQ14an and PTF12dam. There are no pre-peak spectra available for SN\,2007bi, but the earliest spectrum (at phase +54 days) is remarkably similar to the PS1-14bj spectra, with the main difference being that SN\,2007bi has slightly higher photospheric velocities of 12,000~\kms\, \citep{gmo+09}. LSQ14an at +60 days is also very similar to PS1-14bj despite the difference in phase, including the low velocities. This speaks to the very slow spectroscopic evolution of these objects, mirroring the slow light curves. PTF12dam, in contrast, shows a much bluer spectrum with O II features pre-peak, similar to more typical H-poor SLSNe \citep{qkk+11}, consistent with its much higher temperatures near peak.

\begin{figure}
\centering
\includegraphics[width=3.4in]{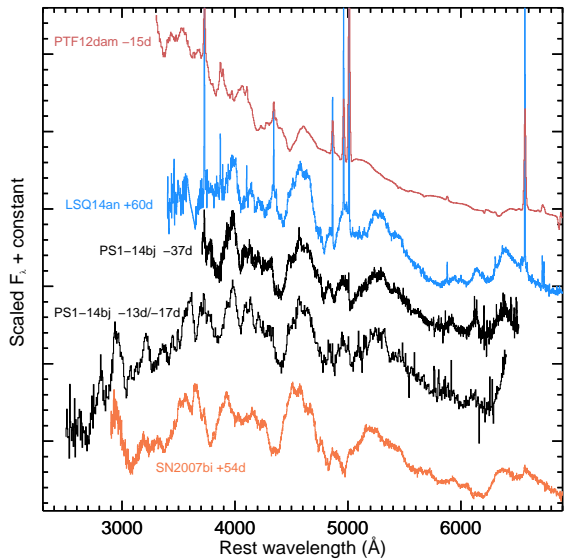}
\caption{Early-time spectra of PS1-14bj compared to other slowly-evolving SLSNe \citep{nsj+13, gmo+09}. The spectra are dominated by features from \ion{Ca}{2}, \ion{Fe}{2} and \ion{Mg}{2}, similar to the earliest available spectra of SN\,2007bi. The early-time spectrum of PTF12dam, in contrast, is dominated by O II features and is much bluer than PS1-14bj at the corresponding phase.  }
\label{fig:earlyspec}
\end{figure}

Late-time spectra are shown in Figure~\ref{fig:latespec}, where we have combined the +155 and +202 days spectra of PS1-14bj for better S/N and wavelength coverage. The spectrum shows a mix of absorption and emission features, and in particular has not turned fully nebular despite the late phase. Such slow evolution is similar to what was seen in SN\,2007bi, which did not turn fully nebular until $> 400$ days past peak. The four SNe shown in Figure~\ref{fig:latespec} show many similarities in their late-time spectra, but with one prominent difference: PS1-14bj and LSQ14an both show strong emission features around 5000\,\AA\, and 4360\,\AA. These features are not present in SN\,2007bi or PTF12dam, and indeed have not been previously seen in SLSN spectra. We tentatively identify these features with [\ion{O}{3}] $\lambda$4363 and $\lambda\lambda$4959,5007. 

A zoom-in of these features is shown in Figure~\ref{fig:oiii}; the FWHM of the line at 4360\,\AA\, is about 3400~\kms\,. The feature around 5000\,\AA\, is broader with a slightly asymmetric profile --- if this is indeed [\ion{O}{3}] emission this can be understood as the blended emission from the stronger line at 5007\,\AA\, and the line at 4959\,\AA. The righthand panel of Figure~\ref{fig:oiii} shows a double-Gaussian fit to the emission near 5000\,\AA, with the line ratios and central wavelengths fixed to those expected for $\lambda\lambda$4959,5007 and with the same FWHM as for the $\lambda$4363 line. This provides a reasonable match to the line profile, although there is excess blue flux compared to the fit particularly in the line at 5007\,\AA\,. 

Broad [\ion{O}{3}] emission is usually not seen in SNe until years to decades following core-collapse, as well as in O-rich supernova remnants (e.g., \citealt{mfc+12}). Instead, the dominant emission from oxygen tends to be [\ion{O}{1}] $\lambda\lambda$6300,6364, which indeed is what was eventually seen in SN\,2007bi and PTF12dam \citep{gmo+09,ysv+10,nsj+13,csj+15}. We discuss possible sources of the ionizing flux in Section~\ref{sec:models}. We note that the low flux ratio of [\ion{O}{3}] $\lambda\lambda$4959,5007 to [\ion{O}{3}] $\lambda$4363 ($\lesssim 3$) implies that the electron density in the [\ion{O}{3}]-emitting region is near the critical density for these transitions ($\gtrsim 10^6~{\rm cm}^{-3}$; e.g. \citealt{fgf+99}). A similar [\ion{O}{3}] line ratio was observed in the Type IIn SN\,1995N, with a derived density in the [\ion{O}{3}] emitting region of $3\times10^8~{\rm cm}^{-3}$ \citep{fcf+02}.

\begin{figure}
\centering
\includegraphics[width=3.4in]{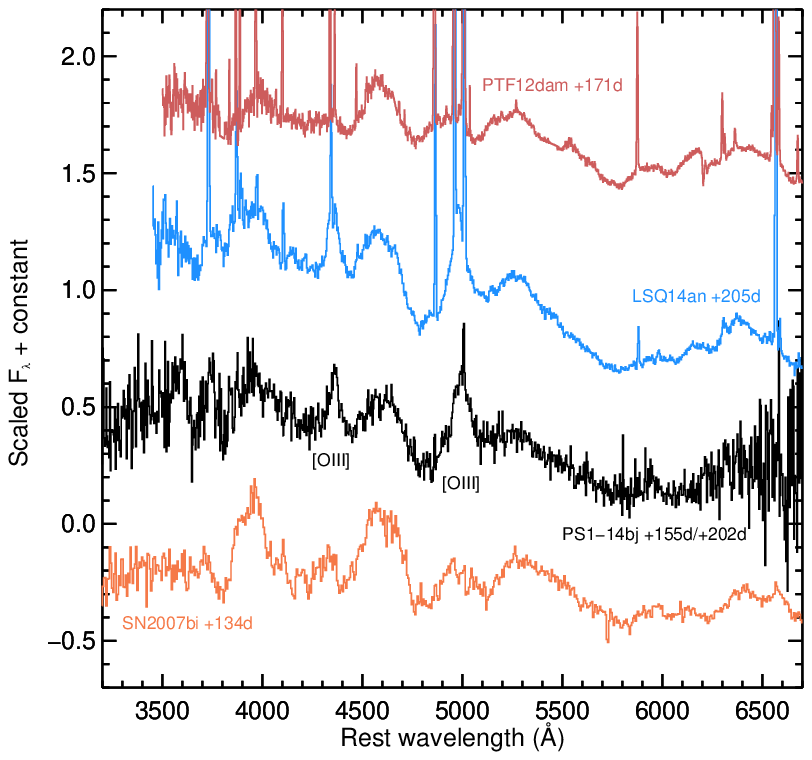}
\caption{Late-time spectrum of PS1-14bj compared to other slowly-evolving SLSNe \citep{nsj+13, ysv+10}. Spectra are labeled with their rest-frame phases relative to maximum light. Notably, PS1-14bj and LSQ14an both show strong emission features near 4360\,\AA\, and 5000\,\AA\, that are not seen in the other SNe, and which we identify as [\ion{O}{3}] emission. Note that these spectra are not host-subtracted, so there is a contribution from the host galaxy to the continuum in the spectra as displayed. The host-subtracted photometry of PS1-14bj shows that its color temperature stays nearly constant around 8000-10,000~K to very late times, however, so not all of the blue continuum can be attributed to the host.}
\label{fig:latespec}
\end{figure}

\begin{figure}
\centering
\includegraphics[width=3.4in]{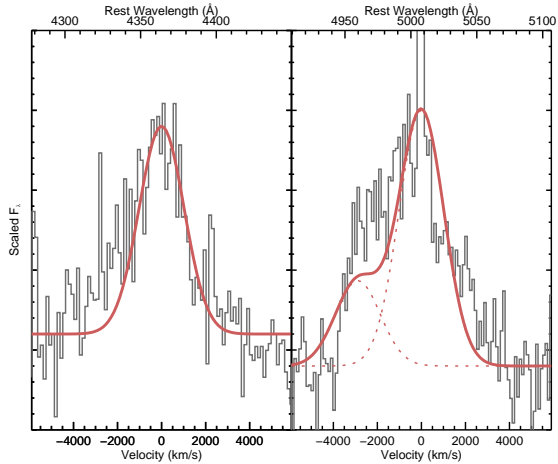}
\caption{Zoom-in on the line profiles of the emission features at 4360\,\AA\ and 5000\,\AA\ in the late-time spectrum of PS1-14bj. The red line shows a Gaussian fit with a FWHM of 3400~\kms\,, assuming the correct identifications for these features are [\ion{O}{3}] $\lambda$4363 and [\ion{O}{3}] $\lambda\lambda$5007,4959.
}
\label{fig:oiii}
\end{figure}

\section{PS1-14bj Host Galaxy}
\label{sec:hostgal}
The host galaxy of PS1-14bj is detected in our deep PS1/MDS pre-explosion stacks (Section~\ref{sec:hostgal_phot}; Figure~\ref{fig:hostgal}) in filters \gps\rps\ips\zps. \rps\ corresponds approximately to rest-frame $B$-band, giving an absolute magnitude $M_B \simeq -18.1~{\rm mag}$. Fitting a galaxy model to the SED using the FAST code \citep{kvl+09}, we find a best-fit galaxy mass of $\log({\rm M_{gal}}/M_{\odot}) = 8.5 \pm 0.5$. Other galaxy properties going into the model grid (such as extinction, stellar population age and star formation rate) are not well constrained by our relatively noisy galaxy photometry. The detection of narrow [\ion{O}{3}] $\lambda$5007 and H$\beta$ emission from the host galaxy in our late time spectra indicates that the host galaxy is star forming. While we do not detect narrow H$\alpha$, we can calculate a lower limit on the star formation rate from the H$\beta$ flux assuming no host galaxy extinction. Using H$\beta$ and the relation from \citet{ken98} we calculate a star formation rate of $\gtrsim 0.2~M_{\odot}~{\rm yr}^{-1}$.

In the context of H-poor SLSNe, the host galaxy of PS1-14bj is quite typical, with both its absolute magnitude and mass close to the median values found in the compilation of \citet{lcb+14} ($\langle M_B \rangle = -17.6~{\rm mag}$ and $\langle {\rm M_{gal}} \rangle = 2 \times 10^8 M_{\odot}$, respectively). While there are relatively few slowly-evolving H-poor SLSNe known, their host galaxies so far do not appear to be different from the bulk of H-poor SLSNe, spanning a similar range of properties \citep{csj+15}; PS1-14bj continues this trend.

\section{Energy Sources}
\label{sec:models}
In this section, we discuss potential energy sources for PS1-14bj, and compare the observed properties to model predictions.

\subsection{Radioactive Decay: PS1-14bj as a PISN}
$^{56}$Ni decay is usually ruled out as the power source of ordinary H-poor SLSNe from the shapes of their light curves, with the combination of high peak luminosities and relatively fast-evolving light curves leading to an unphysical solution with $M_{\rm Ni} \gtrsim M_{\rm ej}$ (e.g., \citealt{ccs+11,lcb+13}). Given the broad light curve and lower peak luminosity of PS1-14bj, combined with the decline rate matching the decay of $^{56}$Co, we explore radioactive decay. 

Figure~\ref{fig:arnett} shows $^{56}$Ni-powered fits to the bolometric light curve of PS1-14bj, following \citet{arn82} and \citet{vbc+08}. This is essentially a scaled-up model of a Type Ic supernova, with the shape of the light curve determined from two main parameters: the total Ni mass, $M_{\rm Ni}$, and the diffusion time through the ejecta $\tau_m \propto \kappa^{1/2} M_{\rm ej}^{3/4} E_{\rm K}^{-1/4}$, where $\kappa$ is the opacity, $M_{\rm ej}$ the total ejecta mass, and $E_{\rm K}$ the kinetic energy. Using the velocity at peak measured from our spectra, $\simeq 5,000~{\rm km s}^{-1}$ and an opacity $\kappa = 0.1~{\rm cm}^2{\rm g}^{-1}$, we derive a total $^{56}$Ni mass $M_{\rm Ni} \sim 10$~M$_{\odot}$ and a total ejecta mass $M_{\rm ej} \sim 52$~M$_{\odot}$ for the best-fit model, which has a reduced $\chi^2$ value of 1.9. Note that the derived mass here scales inversely with the assumed opacity, which is uncertain by at least a factor of two in either direction (see Appendix D of \citealt{isj+13} for a discussion). As a result, the inferred ejecta masses are also uncertain by at least a factor of two. This caveat applies to any mass estimates also in the following sections.
The Ni-powered  model is a good fit to the data through $\sim 200~{\rm days}$ past peak, but underpredicts the data points measured at $\sim 400~{\rm days}$, suggesting that even if $^{56}$Ni decay powers the main peak a different mechanism is likely dominating at these late times. We caution, however, that the bolometric correction of these last three points is uncertain, so the departure from the $^{56}$Co decay slope may not be significant. The kinetic energy of this model is $2.2 \times 10^{52}~{\rm erg}$, (subject to the same caveats regarding effective opacity, and thus also uncertain by at least a factor of two; see also \citealt{wjc15}). This gives a ratio of radiated to kinetic energy is $\sim 3\%$, of the same order as is seen in Type Ibc SNe. 

If $^{56}$Ni is the power source, the inferred nickel mass and total ejecta mass imply that PS1-14bj likely exploded as a pair-instability SN rather than a core-collapse SN. Figure~\ref{fig:pisn_comp} shows the light curve of PS1-14bj compared to the theoretical PISN light curves from \citet{kwh11}, \citet{dwl+13} and \citet{kbl+14}. Overall, PS1-14bj is a reasonable match to the shape of these light curves, with a luminosity intermediate between the He100 and He110 models. It does have a flatter light curve peak and lower peak luminosity than the similar-width PISN light curves, however. Unlike the simple Arnett model fit in Figure~\ref{fig:arnett}, these light curves are derived from radiation hydrodynamics modeling of the supernova explosions given several different progenitor models, and therefore do not have the freedom to tune the nickel and ejecta mass separately as in the simple Arnett fit. This explains why the fit in Figure~\ref{fig:arnett} can better reproduce the flatter light curve shape, even though both the PISN models and the Arnett model are essentially Ni-powered. The PISN models have the same problem as the Arnett model in that the light curve points around $\sim 400~{\rm days}$ lie above the model prediction. We note that while not a perfect match to the theoretical expectation, PS1-14bj at least shows that there exist SLSNe with the broad light curves and long rise times expected by PISN models.

\citet{mtt+10} presented core-collapse models for the light curve of SN\,2007bi, showing that it could also be reasonably fit as an extreme core-collapse event rather than a PISN. In their model, the light curve is fit by the explosion of a 43~M${\odot}$ carbon-oxygen core, producing 6~M$_{\odot}$ of $^{56}$Ni, and thus demonstrating that such large Ni masses need not be the result of a PISN. The main discriminant between the core-collapse and PISN models is the supernova rise time, with the core-collapse having a lower ejected mass and thus a shorter rise time. PS1-14bj is not a good match to their core-collapse model, however: its rise time of $\gtrsim 125$~days, as well as the relatively low velocities on the rise, are better matched by their PISN model. Thus, if PS1-14bj is powered by $^{56}$Ni it is more consistent with a PISN than an extreme core-collapse SN interpretation.

Arguing against a PISN/$^{56}$Ni model of PS1-14bj, however, are the color and spectroscopic evolution. While PS1-14bj is redder than a typical SLSN on the rise, it is still bluer than the expected colors of a PISN explosion due to the large amount of line blanketing from $^{56}$Ni (e.g., the models of \citealt{dhw+12,dwl+13} have a maximum photospheric temperature of $\sim 6000~{\rm K}$). The discrepancy grows worse after peak, as PISN models (or indeed any $^{56}$Ni-powered model) predict the color turning redder on the light curve decline, whereas PS1-14bj shows a nearly constant color that, if anything, turns bluer. We note that the 250M model of \citet{kbl+14} shows a temperature plateau for about 200 days up to and around peak light, but even this model cools to a color temperature of $\sim 5000~{\rm K}$ on the decline.

The spectral evolution of PS1-14bj also does not match what is expected from a PISN model. \citet{kwh11} found a reasonable match between the spectra of their He100 model at 100 days after explosion with the earliest available spectra of SN~2007bi, despite not explicitly tuning the model.  Given the similarity between the two objects, that model should also be a reasonable match to the spectra of PS1-14bj near maximum light, except for the difference in expansion velocities. However, the spectra of the He100 model show significant evolution over time, with a smooth and relatively blue spectrum at 50 days after explosion evolving to a very red and line-blanketed one at 250 days (their Figure 9; \citealt{kwh11}).  The spectra of PS1-14bj do not show such a rapid evolution and the models would not be a good match to our earliest and latest spectra (Figure~\ref{fig:spec_montage}).
Recently, \citet{jsh16} calculated model nebular phase spectra of PISNe, which are also not a good match to our late-time spectra. Specifically, their models have very little flux emerging bluewards of 5000~\AA\, and are dominated by lines from \ion{Fe}{2} in the wavelength range covered by our spectra. Some contributing flux from [\ion{O}{1}] $\lambda\lambda$6300,6364 is seen in their models, but no higher-ionization species of oxygen are predicted. Beyond the theoretical models, the fact that [\ion{O}{3}] is not seen in nebular spectra of ordinary core-collapse SNe is in itself an argument against $^{56}$Ni decay being the sole power source.

\begin{figure}
\centering
\includegraphics[width=3.4in]{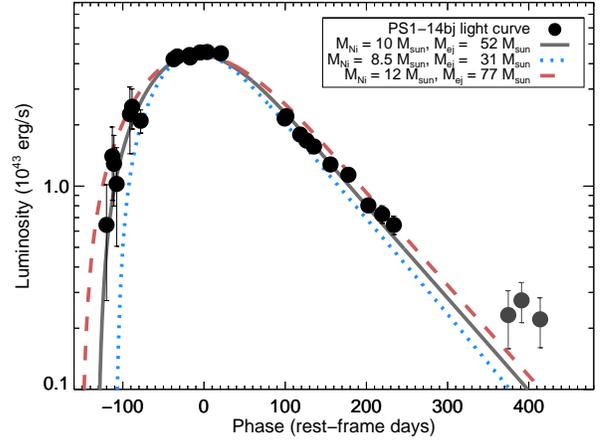}
\caption{Fits to the light curve of PS1-14bj, using a Ni-powered model following \citet{arn82}. The observed bolometric light curve can be reproduced with a model having a total $^{56}$Ni mass of 10~M$_{\odot}$, and a total ejecta mass of $\sim 50~{\rm M}_{\odot}$. Note that the bolometric correction of the last three points (colored gray) is highly uncertain, so the departure from the $^{56}$Co decay slope may not be real.}
\label{fig:arnett}
\end{figure}

\begin{figure}
\centering
\includegraphics[width=3.4in]{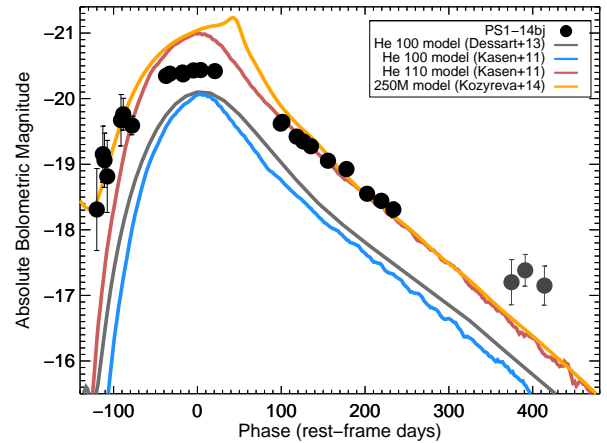}
\caption{PS1-14bj bolometric light curve (black points) compared to theoretical PISN light curves from \citet{kwh11}, \citet{dwl+13} and \citet{kbl+14}. PS1-14bj has a lower peak luminosity compared to its width than the theoretical curves, but matches the overall long timescales well.}
\label{fig:pisn_comp}
\end{figure}

\subsection{Magnetar Models}
Since radioactive decay cannot be the energy source for the majority of SLSNe, several alternatives have been proposed. One possible energy source is the rotational energy of a newborn neutron star rapidly spinning down in a strong magnetic field. Theoretical calculations show that in the regime with initial spin of a few milliseconds and magnetic field of $10^{14}-10^{15}$~G, the magnetar spin-down time is comparable to the diffusion time through the supernova ejecta, producing a long-lived energy injection that can significantly boost the optical radiation \citep{kb10,woo10}. Magnetar models have successfully been able to reproduce the light curves and other basic observed properties of a range of SLSNe (e.g., \citealt{ccs+11,isj+13,lcb+13}), including those with slow decay time scales such as SN\,2007bi \citep{dhw+12, nsj+13}. 

The key parameters of the magnetar model are the initial spin period $P$, which sets the total amount of rotational energy as $E_p \simeq 2 \times 10^{52}~{\rm erg} \times (P/1~{\rm ms})^{-2}$; the magnetic field, which together with the initial period sets the spin-down timescale as  $\tau_p \simeq 4.7~{\rm days} \times (P/1~{\rm ms})^{2} \times (B/10^{14}~{\rm G})^{-2}$, and the total ejecta mass which sets the diffusion timescale $\tau_m$ as before. The resulting luminosity can then be solved for semi-analytically as

\begin{equation}
L(t) = \frac{2 E_p}{\tau_p \tau_m}e^{-(\frac{t}{\tau_m})^2}  
\times \int_0^{t} \frac{1}{(1+t'/\tau_p)^2} e^{(\frac{t'}{\tau_m})^2} \frac{t'}{\tau_m} dt' .
\label{eqn:magnetar}
\end{equation} 

\noindent This suggests that in order to create a magnetar-powered SLSN with slow timescales, we need some combination of a larger ejecta mass (increasing the diffusion time) and larger ratio of $P/B$ (increasing the spin-down timescale) compared to more typical SLSNe. Since $P$ sets the total energy scale, which is not that different in PS1-14bj to other SLSNe, in practice this means we require a weaker magnetic field.

The green-dotted curve in Figure~\ref{fig:magnetar} shows the best-fit three-parameter magnetar model to the light curve of PS1-14bj. This model has an initial spin of 3.1~ms, a magnetic field of $10^{14}$~G, and a total ejecta mass of 22.5~M$_{\odot}$. It does a decent job of reproducing the overall broad light curve, but overpredicts the flux at late times and also starts declining earlier than the observed light curve; the reduced $\chi^2$ of the fit is 2.9. 

Overpredicting the flux on the late light curve decline is a common problem of magnetar models, but can be overcome if the assumption that all of the spin-down energy from the magnetar is being thermalized in the ejecta is relaxed. As the ejecta expand, they become optically thin to hard emission and significant amounts of $\gamma$- and X-ray photons can escape. \citet{wwd+15} considered this effect for SLSNe, and parametrized the changing opacity to $\gamma$-rays as $\tau_{\gamma} = A t^{-2}$, so that Equation~\ref{eqn:magnetar} is modified by the expression $(1-e^{-At^{-2}})$ \citep{cwv09,cwv12}. Including this term, the purple-dashed curve in Figure~\ref{fig:magnetar} shows the best fit to PS1-14bj, found by first finding the best three-parameter fit to the data on the rise through peak, and then adjusting the leakage parameter to fit the decline. The model shown also has an initial spin of 3.1~ms but a lower magnetic field of $5 \times 10^{13}$~G, a slightly lower ejecta mass of 16~M$_{\odot}$, a leakage parameter $A = 5.5 \times 10^{14}~{\rm s}^2$ and a reduced $\chi^2$ value of 1.9. This value of $A$ is within the range of typical values for CCSNe ($10^{13} - 10^{15}~{\rm s}^2$) though lower by a factor of $\sim 5$ than what one would derive from Equation~4 of \citet{wwd+15} using the mass derived in the magnetar model and the velocities we measure from the spectra ($A \propto \kappa_{\gamma} M_{\rm ej} / v^2$), suggesting either a lower effective gamma-ray opacity $\kappa_{\gamma}$, or that the assumptions of uniform ejecta density distribution and constant expansion velocity may not apply.

\begin{figure}
\centering
\includegraphics[width=3.4in]{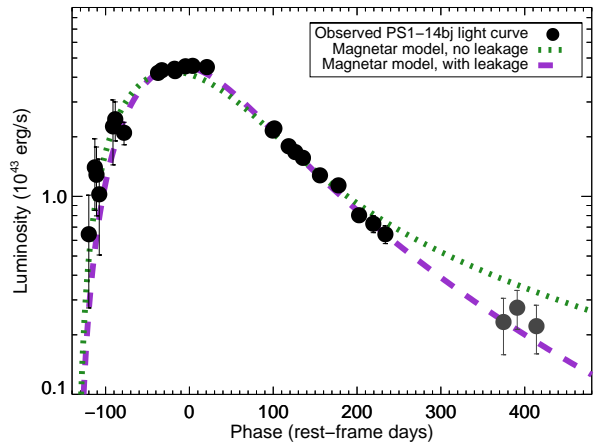}
\caption{Magnetar model fits to the light curve of PS1-14bj. The green dotted curve is the best fit basic magnetar model, while the purple dashed curve is the best-fit model when allowing for late-time hard emission leakage. Without including the leakage term, the luminosity on the decline is overpredicted.
\label{fig:magnetar}}
\end{figure}

While the overall shape of the light curve can be fit by a magnetar model, this class of models also meets a problem in explaining the observed color evolution, particularly at late times. Modeling the temperature evolution on the rise generally requires hydrodynamical simulations outside of the scope of this paper, but on the decline the magnetar model predicts that the ejecta is swept into a shell moving at constant velocity, and we can estimate the temperature by assuming blackbody radiation and a radius expanding linearly with time at the predicted velocity. For either version of the magnetar model this leads to a blackbody temperature decreasing with time, which is not what we observe for PS1-14bj (Figure~\ref{fig:bb}). 

Again, this simple calculation does not account for radiation from the magnetar itself escaping, however; one could imagine a scenario where the late-time heating seen in PS1-14bj is due to such X-ray to UV breakout from the central power source. Similarly, the [\ion{O}{3}] lines could arise from O-rich ejecta being ionized from within becoming visible; we note that ionization calculations of a pulsar wind nebula inside expanding SN ejecta predict [\ion{O}{3}] $\lambda$5007 to be the strongest line in the optical window in the H- and He-free case \citep{cf92}. Although these calculations were done considering the scenario of a much weaker pulsar wind nebula than the magnetar considered here, and considered the line emission at somewhat later times than our observations, it demonstrates that such a late-time optical signature of the central engine is at least plausible.

\subsection{CSM Interaction}
Interaction between the SN ejecta and H-free CSM is a third alternative for powering SLSNe \citep{ci11,gb12,mm12,mbt+13,cwv12}, and has been shown to reproduce the light curves of a wide range of SLSNe, including H-poor ones with slow evolution such as PTF12dam and SN\,2007bi \citep{cwv+13,nsj+14,csj+15}. Light curves from CSM interaction depend on a number of parameters describing the mass, extent and profile of the CSM, and generally requires hydrodynamical modeling outside of the scope of this paper. For an order-of-magnitude estimate of the required CSM mass, we consider the simplified case of shock breakout from a dense wind described in \citet{ci11}. Here, the CSM profile is described as a wind with $\rho_{\rm CSM} = D r^{-2}$ ($= 5 \times 10^{16} D_* r^{-2} $ in cgs units) out to some radius $R_w$, where $D_*$ is a constant. The diffusion time $t_d$ scales with this density parameter as $t_d = 19.4 \kappa D_*$~(days), where $\kappa$ again is the opacity. Taking the observed rise time ($\sim 128$~days) as an estimate of $t_d$, for an opacity $\kappa = 0.1~{\rm cm}^2~{\rm g}^{-1}$ as in the other models, we get a density parameter $D_* \simeq 66$. The wind radius $R_w$ is approximately the blackbody radius at peak ($\sim 2.5\times 10^{15}$~cm); integrating the density profile we get a required wind mass of $\simeq 50~{\rm M}_{\odot}$. Again, the derived mass scales inversely with the assumed opacity; if we instead use $\kappa = 0.2~{\rm cm}^2~{\rm g}^{-1}$ (appropriate for a fully ionized metal-dominated gas), the derived mass goes down by a factor of two.

This number is subject to a lot of simplifying assumptions, but serves to illustrate that powering a light curve as luminous and as slow as PS1-14bj's with CSM interaction will require a CSM mass of tens of solar masses.  
Alternatively, the unusually long risetime of PS1-14bj might also be accommodated by invoking an unusual CSM density profile, as was necessary to explain the Type IIn SN\,2008iy, which had the longest measured risetime of a supernova known to date ($>400$~d; \citealt{msb+10}).

Currently available CSM interaction models (e.g., \citealt{dah15}) do not produce spectra that have the broad P-Cygni features exhibited by PS1-14bj on the rise and near maximum light (Figure~\ref{fig:earlyspec}).  Instead, the computed spectra tend to exhibit strong emission lines on a continuum with only narrow P-Cygni features blueshifted by the wind velocity.  However, only the hydrogen-rich case of SLSN-II, analogous to normal SNe IIn, has been explored in any detail.
We do have spectra both at peak (+22 days) and at late times (+202 days) that go sufficiently red to cover H$\alpha$, and we do not detect H$\alpha$ emission at either epoch --- this indicates that the CSM material around PS1-14bj would have to be H-poor.

Rather than being the primary power source for the light curve, CSM interaction could also come into play at later times as the expanding SN ejecta encounter a previously ejected shell of material. Evidence for delayed interaction with a CSM shell was recently seen in the SLSN iPTF13ehe \citep{yqo+15}, which was originally classified as a SN\,2007bi-like slowly-evolving SLSN, but developed broad H$\alpha$ emission in its late-time spectra. This emission was interpreted to originate from interaction between the ejecta and a H-rich shell located at a radius $\sim 4 \times 10^{16}~{\rm cm}$, and from the PTF sample of SLSNe with late-time spectra they estimate that such interaction may be present in at least 10-15\% of ``H-poor'' SLSNe. In the Type Ic SN\,2010mb, \citet{bgm+14} argued that interaction with $\sim 3$~M$_{\odot}$ of O-rich material was responsible for the slowly-declining light curve, blue continuum and strong [\ion{O}{1}] $\lambda$5577 emission seen at late times.

Similarly, CSM interaction could be the source of the persistent blue continuum and [\ion{O}{3}] lines seen in PS1-14bj. In particular, broad [\ion{O}{3}] emission has been seen in several core-collapse SNe with interaction in years to decades after explosion and in SN remnants \citep{fgf+99,mf08}, and interpreted as O-rich ejecta being ionized by the reverse shock. A similar feature was recently seen in the Type Ib SN\,2014C at a phase of $+282$~days past maximum light, also accompanied by H$\alpha$, X-ray and radio emission all consistent with CSM interaction \citep{mmk+15,mkm+16}. We note that the line widths of 3000-4000~\kms\, seen in PS1-14bj suggest that the [\ion{O}{3}] emission is originating in the ejecta rather than in shocked CSM, likely arising in the freely expanding ejecta being illuminated by the X-rays from the reverse shock \citep{fra84, cf94}. Alternatively, if the [\ion{O}{3}] lines are instead originating in an oxygen-dominated CSM shell, this shell would have to be moving at higher velocities than are typically seen in CSM interaction in order to explain the line widths.

In the scenario where CSM interaction is responsible for the late-time heating of PS1-14bj, iPTF13ehe \citep{yqo+15} makes for an interesting comparison. In particular, while the broad H$\alpha$ feature in iPTF13ehe was interpreted as a result of interaction, iPTF13ehe did not show any corresponding [\ion{O}{3}] features like we see in PS1-14bj. At face value this indicates that the temperature and density conditions in the oxygen-emitting region would be different between the two objects. We note, however, that in other SNe with interaction signatures and late-time [\ion{O}{3}] emission, the [\ion{O}{3}] features are much weaker than H$\alpha$ \citep{fgf+99,mmk+15}. If the flux ratio of H$\alpha$ to [\ion{O}{3}] was similar in iPTF13ehe to these previous objects, we would not expect to see it above the noise level even if present. Therefore, while we cannot say much quantitatively about the ionization conditions in iPTF13ehe versus PS1-14bj, they both illustrate the importance of late-time spectra in shedding light on the nature of SLSN-I.

If we assume that the [\ion{O}{3}] lines in PS1-14bj are arising as a result of CSM interaction, we can estimate the distance to the shell based on the time at which the features emerge and the velocities of the SN ejecta. The minimum value for the forward shock velocity is the maximum velocity in our fit to the spectrum ($\simeq 10,000$~\kms\,), and the broad [\ion{O}{3}] line is detected in our $+137~{\rm day}$ spectrum, corresponding to $\sim 265$~days after our estimated explosion date. This yields a shell distance of $\sim 2 \times 10^{16}~{\rm cm}$, similar to what was derived for iPTF13ehe \citep{yqo+15}.

\section{Conclusions}
\label{sec:conc}
We have shown that PS1-14bj is an unusually slowly evolving SLSN with a number of peculiar properties. Its key observed features can be summarized as follows:
\begin{itemize}
    \item An exceptionally slow rise to maximum light, observed to be $\gtrsim 125$~days, the longest yet reported in the literature for a H-poor SLSN. Independent of model, this implies a large ejecta mass.
    \item Continued slow evolution, with a nearly-flat peak and exponential decline over the first $\sim 250$~days past peak at a best-fit rate of $(1.00 \pm 0.05) \times 10^{-2}~{\rm mag~day}^{-1}$, a remarkably close match to fully-trapped $^{56}$Co decay.
    \item A peak bolometric luminosity of $4.6 \times 10^{43}$~erg/s, and an estimated total radiated energy of $\gtrsim 7.6 \times 10^{50}$~erg.
    \item Unusual color evolution, with the color temperature rising prior to peak, and staying constant within our uncertainties around 8,000-10,000~K through the peak and decline.
    \item A spectrum near peak dominated by features of \ion{Mg}{2}, \ion{Ca}{2} and \ion{Fe}{2}, 
    similar to other slowly-evolving SLSNe including SN\,2007bi, although with lower velocities. 
    \item A late-time spectrum (150-200 days past peak) with strong emission features which we identify as [\ion{O}{3}] $\lambda$4363 and [\ion{O}{3}] $\lambda\lambda$4959,5007, with a velocity width of 3000-4000~\kms\,.
\end{itemize}

All of these properties taken together are not easily explained by any of the single models suggested to power SLSNe. The long rise time, initially red colors, and decline rate initially following $^{56}$Co decay would support a PISN interpretation.
In the classification scheme of \citet{gal12}, PS1-14bj is similar to other objects identified as SLSN-R, such as the prototype SN\,2007bi \citep{gmo+09,ysv+10}.  However, that scheme assumes that the underlying power source is radioactive decay to separate SLSN-R from other SLSN-I.  Here, we have shown that a model powered purely by cobalt decay cannot match the colors or the late-time observations:
PISN models are considerably redder than what is observed for PS1-14bj past peak, and predict a very different late-time spectrum than what we observe. We note that PS1-14bj at least demonstrates that there exist SLSNe with the long rise times predicted by pair-instability models, which has not been observed before. By extension, it is possible that SN\,2007bi may also have had a long rise-time.

A simple magnetar model can reproduce the long time scales by a combination of lower magnetic field and larger ejecta mass than is typically used to fit SLSNe. However, this model has the same problem as the Ni-powered model in explaining the near-constant rather than reddening color past peak, and generally overpredicts the late-time luminosity ($\gtrsim 150$~days past peak). Both of these shortcomings could be overcome by relaxing the assumption that all the radiation from the magnetar is being thermalized in the ejecta; in this scenario the late time heating would be due to emission from the pulsar wind nebula breaking out, and the [\ion{O}{3}] lines similarly coming from inner parts of the ejecta being ionized by the central source. 

Alternatively, the persistent blue continuum and [\ion{O}{3}] lines could arise from material being shocked and ionized by interaction with a CSM shell. Such lines are often seen in core-collapse SNe with CSM interaction, though typically at later phases than we observe here. This scenario could be distinguished from the magnetar breakout by continued spectral monitoring as the line velocities are expected to decrease in the CSM scenario, and stay constant or even increase in the magnetar case. X-ray or radio observations are another possibility.

PS1-14bj further demonstrates the need to explore hybrid models that combine multiple power sources contributing at different points in the light curve (e.g., \citealt{cwv12}). For example, \citet{yqo+15} proposed that interaction with CSM shells emitted during the pulsational pair instability \citep{wbh07} could explain the late-time H emission of iPTF13ehe. Our discovery of emerging [\ion{O}{3}] features in the late-time spectrum further demonstrates the potential of late-time monitoring in revealing the nature of SLSNe. The discovery of a second SLSN, LSQ14an, with late-time features similar to PS1-14bj in its spectrum shows that they are not unique, and may not be rare in slowly-evolving SLSNe.

\acknowledgments
We thank Dan Kasen and Alexandra Kozyreva for providing their theoretical PISN light curves, Andy Monson for help with processing the FourStar data, and Robert Quimby, Dan Perley, Matt Nicholl, Avishay Gal-Yam and Lin Yan for helpful discussions. R. Lunnan acknowledges helpful interactions with Dan Kasen, Lars Bildsten and Eliot Quataert at a PTF Theory Network Retreat, that was funded by the Gordon and Betty Moore Foundation through Grant GBMF5076.

The Pan-STARRS1 Surveys (PS1) have been 
made possible through contributions by the Institute for Astronomy, the 
University of Hawaii, the Pan-STARRS Project Office, the Max-Planck 
Society and its participating institutes, the Max Planck Institute for 
Astronomy, Heidelberg and the Max Planck Institute for Extraterrestrial 
Physics, Garching, The Johns Hopkins University, Durham University, 
the University of Edinburgh, the Queen's University Belfast, the 
Harvard-Smithsonian Center for Astrophysics, the Las 
Cumbres Observatory Global Telescope Network Incorporated, the 
National Central University of Taiwan, the Space Telescope Science Institute, and the National 
Aeronautics and Space Administration under Grant No. NNX08AR22G issued 
through the Planetary Science Division of the NASA Science Mission 
Directorate, the National Science Foundation Grant No. AST-1238877,
the University of Maryland, Eotvos Lorand University (ELTE),
and the Los Alamos National Laboratory. 

Based in part on observations obtained under PIDs GN-2015A-FT-1 (PI: R. Lunnan) and GN-2014A-Q-76/GS-2014A-Q-63 (PI: R. Chornock) at the Gemini Observatory, which is operated by the 
Association of Universities for Research in Astronomy, Inc., under a cooperative agreement 
with the NSF on behalf of the Gemini partnership: the National Science Foundation 
(United States), the National Research Council (Canada), CONICYT (Chile), the Australian 
Research Council (Australia), Minist\'{e}rio da Ci\^{e}ncia, Tecnologia e Inova\c{c}\~{a}o 
(Brazil) and Ministerio de Ciencia, Tecnolog\'{i}a e Innovaci\'{o}n Productiva (Argentina).
This paper includes data gathered with the 6.5 meter Magellan Telescopes located at Las Campanas Observatory, Chile.
This work was supported in part by the Fermi National Accelerator Laboratory which is operated by Fermi Research Alliance, LLC under Contract No. DE-AC02-07CH11359 with the United States Department of Energy.
This work is based in part on observations obtained at the MDM Observatory, operated by Dartmouth College, Columbia University, Ohio State University, Ohio University, and the University of Michigan.
This work was supported in part by the GROWTH project funded by the National Science Foundation under Grant No 1545949. 
This work was supported in part by National Science Foundation Grant No. PHYS-1066293 and the hospitality of the Aspen Center for Physics.
The CfA Supernova Program is supported by NSF grants AST−1211196 and AST−156854 to the Harvard College Observatory. 
SJS acknowledges funding from the European Research Council under the European Union's Seventh Framework Programme (FP7/2007-2013)/ERC Grant agreement n$^{\rm o}$ [291222] and STFC grants ST/I001123/1 and ST/L000709/1. 
Some of the computations in this paper were run on the Odyssey cluster supported by the FAS Division of Science, Research Computing Group at Harvard University. 
This research has made use of NASA's Astrophysics Data System.

\textit{Facilities:} \facility{PS1}, \facility{MMT}, \facility{Magellan:Baade}, \facility{Magellan:Clay}, \facility{Gemini:Gillett}, \facility{Gemini:South},
\facility{LBT}, \facility{McGraw-Hill}, \facility{Hiltner}

\clearpage
\newpage
\begin{deluxetable}{lcccc}
\tablewidth{0pt}
\tabletypesize{\scriptsize}
\tablecaption{PS1-14bj Photometry}
\tablehead{
\colhead{MJD} & 
\colhead{Rest-frame Phase} & 
\colhead{Filter} &
\colhead{AB Mag} &
\colhead{Instrument} \\
\colhead{(days)} &
\colhead{(days)}  & 
\colhead{} &
\colhead{} &
\colhead{} 
}
\startdata
56709.3 & $-60.5$ & \gps & $22.70 \pm 0.40$ & PS1 \\
56751.2 & $-32.9$ & $g$ & $22.34 \pm 0.06$ & MMT \\
56774.3 & $-17.7$ & $g$ & $22.43 \pm 0.03$ & GN \\
56775.3 & $-17.1$ & $g$ & $22.59 \pm 0.03$ & GN \\
56836.0 & $22.8$ & $g$ & $22.52 \pm 0.03$ & LDSS \\
56952.5 & $99.4$ & $g$ & $22.97 \pm 0.05$ & LBT \\
56981.4 & $118.4$ & $g$ & $23.17 \pm 0.05$ & MMT \\
56993.5 & $126.3$ & $g$ & $23.29 \pm 0.05$ & MMT \\
57038.2 & $155.7$ & $g$ & $23.45 \pm 0.06$ & IMACS \\
57071.9 & $177.8$ & $g$ & $23.66 \pm 0.08$ & IMACS \\
57109.3 & $202.4$ & $g$ & $24.16 \pm 0.15$ & GN \\
57135.0 & $219.4$ & $g$ & $24.00 \pm 0.10$ & IMACS \\
57158.0 & $234.5$ & $g$ & $24.06 \pm 0.11$ & IMACS \\
56636.6 & $-108.2$ & \rps & $23.25 \pm 0.32$ & PS1 \\
56666.6 & $-88.5$ & \rps & $22.83 \pm 0.26$ & PS1 \\
56682.6 & $-78.0$ & \rps & $22.39 \pm 0.11$ & PS1 \\
56744.3 & $-37.4$ & $r$ & $21.58 \pm 0.02$ & GN \\
56751.2 & $-32.9$ & $r$ & $21.50 \pm 0.03$ & MMT \\
56774.3 & $-17.7$ & $r$ & $21.50 \pm 0.02$ & GN \\
56775.3 & $-17.1$ & $r$ & $21.54 \pm 0.03$ & GN \\
56794.2 & $-4.7$ & $r$ & $21.47 \pm 0.03$ & MMT \\
56807.1 & $3.8$ & $r$ & $21.48 \pm 0.03$ & MMT \\
56833.0 & $20.8$ & $r$ & $21.53 \pm 0.02$ & LDSS \\
56952.5 & $99.4$ & $r$ & $22.18 \pm 0.03$ & LBT \\
56955.5 & $101.4$ & $r$ & $22.31 \pm 0.06$ & MDM \\
56981.5 & $118.4$ & $r$ & $22.52 \pm 0.04$ & MMT \\
56993.4 & $126.3$ & $r$ & $22.57 \pm 0.04$ & MMT \\
57007.3 & $135.4$ & $r$ & $22.60 \pm 0.05$ & IMACS \\
57038.2 & $155.7$ & $r$ & $22.74 \pm 0.04$ & IMACS \\
57071.9 & $177.8$ & $r$ & $22.89 \pm 0.06$ & IMACS \\
57109.3 & $202.4$ & $r$ & $23.08 \pm 0.05$ & GN \\
57135.1 & $219.4$ & $r$ & $23.23 \pm 0.10$ & IMACS \\
57157.0 & $233.8$ & $r$ & $23.31 \pm 0.06$ & IMACS \\
57371.4 & $374.7$ & $r$ & $24.72 \pm 0.20$ & IMACS \\
57396.3 & $391.0$ & $r$ & $24.45 \pm 0.14$ & IMACS \\
57431.2 & $414.0$ & $r$ & $24.53 \pm 0.17$ & IMACS \\
56656.6 & $-95.1$ & \ips & $22.70 \pm 0.17$ & PS1 \\
56661.6 & $-91.8$ & \ips & $22.52 \pm 0.11$ & PS1 \\
56668.5 & $-87.3$ & \ips & $22.42 \pm 0.21$ & PS1 \\
56683.6 & $-77.4$ & \ips & $22.01 \pm 0.08$ & PS1 \\
56744.3 & $-37.4$ & $i$ & $21.35 \pm 0.02$ & GN \\
56751.2 & $-32.9$ & $i$ & $21.34 \pm 0.02$ & MMT \\
56774.3 & $-17.7$ & $i$ & $21.35 \pm 0.02$ & GN \\
56775.3 & $-17.1$ & $i$ & $21.28 \pm 0.02$ & GN \\
56807.2 & $3.8$ & $i$ & $21.19 \pm 0.02$ & MMT \\
56833.0 & $20.8$ & $i$ & $21.37 \pm 0.02$ & LDSS \\
56952.5 & $99.4$ & $i$ & $22.14 \pm 0.03$ & LBT \\
56981.5 & $118.4$ & $i$ & $22.37 \pm 0.05$ & MMT \\
56993.4 & $126.3$ & $i$ & $22.35 \pm 0.05$ & MMT \\
57007.3 & $135.4$ & $i$ & $22.47 \pm 0.03$ & IMACS \\
57038.3 & $155.7$ & $i$ & $22.79 \pm 0.04$ & IMACS \\
57071.9 & $177.8$ & $i$ & $22.85 \pm 0.08$ & IMACS \\
57109.3 & $202.4$ & $i$ & $22.92 \pm 0.06$ & GN \\
57129.1 & $215.5$ & $i$ & $23.05 \pm 0.05$ & MDM \\
57131.2 & $216.9$ & $i$ & $23.21 \pm 0.10$ & MMT \\
57135.1 & $219.4$ & $i$ & $23.16 \pm 0.08$ & IMACS \\
57157.0 & $233.8$ & $i$ & $23.41 \pm 0.10$ & IMACS \\
57371.3 & $374.6$ & $i$ & $24.24 \pm 0.15$ & IMACS \\
57396.3 & $391.0$ & $i$ & $24.12 \pm 0.12$ & IMACS \\
57431.2 & $414.0$ & $i$ & $24.48 \pm 0.20$ & IMACS \\
56618.6 & $-120.1$ & \zps & $23.53 \pm 0.53$ & PS1 \\
56629.6 & $-112.8$ & \zps & $22.69 \pm 0.27$ & PS1 \\
56632.6 & $-110.9$ & \zps & $22.78 \pm 0.25$ & PS1 \\
56637.6 & $-107.6$ & \zps & $23.03 \pm 0.44$ & PS1 \\
56662.6 & $-91.2$ & \zps& $22.17 \pm 0.21$ & PS1 \\
56667.7 & $-87.8$ & \zps & $22.08 \pm 0.21$ & PS1 \\
56681.5 & $-78.7$ & \zps & $22.24 \pm 0.19$ & PS1 \\
56774.3 & $-17.7$ & $z$ & $21.71 \pm 0.04$ & GN \\
56775.3 & $-17.1$ & $z$ & $21.78 \pm 0.05$ & GN \\
56833.0 & $20.8$ & $z$ & $21.64 \pm 0.06$ & LDSS \\
56952.5 & $99.4$ & $z$ & $22.60 \pm 0.09$ & LBT \\
57009.3 & $136.7$ & $z$ & $22.99 \pm 0.10$ & IMACS \\
57039.3 & $156.4$ & $z$ & $23.16 \pm 0.11$ & IMACS \\
57071.9 & $177.8$ & $z$ & $23.26 \pm 0.13$ & IMACS \\
57135.1 & $219.4$ & $z$ & $24.00 \pm 0.24$ & IMACS \\
57157.0 & $233.8$ & $z$ & $24.16 \pm 0.27$ & IMACS \\
56757.0 & -29.1 & $J$ & 21.56 $\pm$ 0.10 & FourStar \\
57111.1 & 203.6 & $J$ & 23.24 $\pm$ 0.15 & FourStar 
\enddata
\label{tab:phot}
\end{deluxetable}

\begin{deluxetable}{lcccccccc}
\tablecaption{Log of PS1-14bj Spectroscopic Observations}
\tablehead{
\colhead{UT Date} &
\colhead{Phase\tablenotemark{a}}  &
\colhead{Instrument} &
\colhead{Grating} &
\colhead{Filter}  &
\colhead{Wavelength Range} &
\colhead{Resolution} &
\colhead{Exposure Time} &
\colhead{Mean Airmass} \\
\colhead{} &
\colhead{(days)} &
\colhead{} &
\colhead{} &
\colhead{} &
\colhead{(\AA)} &
\colhead{(\AA)} &
\colhead{(s)} &
\colhead{}
}
\startdata
2014 Mar 08.32 & $-50.6$ & BlueChannel & 300GPM   & None  & 3330$-$8550 & 5.6 & 3600 & 1.21 \\
2014 Mar 28.36   & $-37.4$ & GMOS-N      & R400     & OG515 & 5640$-$9910 & 7.1 & 3600 & 1.06 \\
2014 Apr 27.34   & $-17.7$ & GMOS-N   & R400     & OG515 & 5485$-$9750 & 7.1 & 2435 & 1.26 \\
2014 Apr 28.28   & $-17.1$ & GMOS-N   & R400     & OG515 & 5485$-$9750 & 7.1 & 1800 & 1.07 \\
2014 Apr 28.20   & $-17.2$ & BlueChannel & 300GPM   & None  & 3330$-$8540 & 7.6 & 8000 & 1.29 \\
2014 May 03.99   & $-13.4$ & GMOS-S      & B600     & None  & 3600$-$6430 & 4.6 & 3600 & 1.21 \\
2014 Jun 25.99   & $+21.5$ & LDSS3C       & VPH-All  & None  & 4000$-$10000 & 8.1 & 1500 & 1.84 \\
2014 Jun 27.99   & $+22.8$ & LDSS3C   & VPH-Red  & OG590 & 5925$-$10600 & 4.9 & 3600 & 2.01 \\
2014 Dec 19.30   & $+137.4$ & IMACS       & 200/+15   & None  & 4500$-$10000 & 7.1 & 7200 & 1.33 \\
2015 Jan 15.27   & $+155.1$ & IMACS       & 300/+17.5 & None  & 4200$-$9445 & 4.1 & 3600 & 1.21 \\
2015 Mar 28.32   & $+202.4$ & GMOS-N      & R400     & OG515 & 5980$-$10315 & 7.1 & 7200 & 1.06
\enddata
\label{tab:spec}
\tablenotetext{a}{Phase is in rest-frame days relative to bolometric maximum light.}
\end{deluxetable}

\begin{deluxetable}{lccccccc}
\tablecaption{Log of LSQ14an Spectroscopic Observations}
\tablehead{
\colhead{UT Date} &
\colhead{Instrument} &
\colhead{Grating} &
\colhead{Filter}  &
\colhead{Wavelength Range} &
\colhead{Resolution} &
\colhead{Exposure Time} &
\colhead{Mean Airmass} \\
\colhead{} &
\colhead{} &
\colhead{} &
\colhead{} &
\colhead{(\AA)} &
\colhead{(\AA)} &
\colhead{(s)} &
\colhead{}
}
\startdata
2014 Jan 08.336    & IMACS       & 300/+17.5 & None  &
3700$-$9450 & 4.9 & 2400 & 1.17 \\
2014 Feb 28.375    & IMACS       & 300/+17.5 & None  &
3700$-$10300 & 3.6 & 1800 & 1.11 \\
2014 Jun 25.034   & LDSS3C     & VPH-All  & None  &
4000$-$10000 & 8.1 & 3000 & 1.06 \\
2015  Feb 19.254   & IMACS       & 300/+17.5 & None  &
3700$-$9460 & 4.9 & 1800 & 1.08 \\
2015 Apr 23.300   & IMACS       & 300/+17.5 & None  &
3700$-$9435 & 4.9 & 3000 & 1.49 \\
2015 Jul  17.041  & LDSS3C     & VPH-All  & None  &
4000$-$10000 & 8.1 & 3600 & 1.25
\enddata
\label{tab:lsqspec}
\end{deluxetable}

\begin{deluxetable}{lc}
\tablecaption{PS1-14bj Bolometric Light Curve}
\tablehead{
\colhead{Phase} &
\colhead{Luminosity} \\
\colhead{(rest frame days)} &
\colhead{($10^{43}~{\rm erg s}^{-1}$)}
}
\startdata
$ -120.1$ & $0.64 \pm 0.37$ \\
$ -112.8$ & $1.40 \pm 0.55$ \\
$ -110.9$ & $1.29 \pm 0.49$ \\
$ -107.6$ & $1.02 \pm 0.52$ \\
$ -91.2$ & $2.26 \pm 0.81$ \\
$ -88.5$ & $2.45 \pm 0.55$ \\
$ -78.0$ & $2.10 \pm 0.27$ \\
$ -37.4$ & $4.21 \pm 0.12$ \\
$ -32.9$ & $4.34 \pm 0.13$ \\
$ -17.7$ & $4.41 \pm 0.10$ \\
$ -17.1$ & $4.29 \pm 0.11$ \\
$ -4.7$ & $4.54 \pm 0.13$ \\
$ 3.8$ & $4.56 \pm 0.14$ \\
$ 20.8$ & $4.50 \pm 0.14$ \\
$ 99.4$ & $2.15 \pm 0.09$ \\
$ 101.4$ & $2.21 \pm 0.08$ \\
$ 118.4$ & $1.79 \pm 0.08$ \\
$ 126.3$ & $1.67 \pm 0.07$ \\
$ 135.4$ & $1.57 \pm 0.05$ \\
$ 155.7$ & $1.28 \pm 0.06$ \\
$ 177.8$ & $1.14 \pm 0.08$ \\
$ 202.4$ & $0.80 \pm 0.06$ \\
$ 219.4$ & $0.73 \pm 0.07$ \\
$ 233.8$ & $0.64 \pm 0.07$ \\
$ 374.7$ & $0.23 \pm 0.07$ \\
$ 391.0$ & $0.27 \pm 0.06$ \\
$ 414.0$ & $0.22 \pm 0.06$ 
\enddata
\label{tab:bollc}
\end{deluxetable}

\end{document}